\documentclass[amsmath,showpacs,nofootinbib,11pt]{revtex4-1}
\usepackage{graphicx}
\usepackage{dcolumn}
\usepackage{bm}
\usepackage{color} 
\usepackage{slashed}
\DeclareUnicodeCharacter{2212}{-}

\begin{document}
\newcommand{\hs}{\hspace*{0.5cm}}
\newcommand{\vs}{\vspace*{0.5cm}}
\newcommand{\be}{\begin{equation}}
\newcommand{\ee}{\end{equation}}
\newcommand{\bea}{\begin{eqnarray}}
\newcommand{\eea}{\end{eqnarray}}
\newcommand{\ben}{\begin{enumerate}}
\newcommand{\een}{\end{enumerate}}
\newcommand{\bde}{\begin{widetext}}
\newcommand{\ede}{\end{widetext}}
\newcommand{\nn}{\nonumber}
\newcommand{\crn}{\nonumber \\}
\newcommand{\Tr}{\mathrm{Tr}}
\newcommand{\non}{\nonumber}
\newcommand{\noi}{\noindent}
\newcommand{\al}{\alpha}
\newcommand{\la}{\lambda}
\newcommand{\bet}{\beta}
\newcommand{\ga}{\gamma}
\newcommand{\va}{\varphi}
\newcommand{\om}{\omega}
\newcommand{\pa}{\partial}
\newcommand{\+}{\dagger}
\newcommand{\fr}{\frac}
\newcommand{\bc}{\begin{center}}
\newcommand{\ec}{\end{center}}
\newcommand{\Ga}{\Gamma}
\newcommand{\de}{\delta}
\newcommand{\De}{\Delta}
\newcommand{\ep}{\epsilon}
\newcommand{\varep}{\varepsilon}
\newcommand{\ka}{\kappa}
\newcommand{\La}{\Lambda}
\newcommand{\si}{\sigma}
\newcommand{\Si}{\Sigma}
\newcommand{\ta}{\tau}
\newcommand{\up}{\upsilon}
\newcommand{\Up}{\Upsilon}
\newcommand{\ze}{\zeta}
\newcommand{\ps}{\psi}
\newcommand{\Ps}{\Psi}
\newcommand{\ph}{\phi}
\newcommand{\vph}{\varphi}
\newcommand{\Ph}{\Phi}
\newcommand{\Om}{\Omega}
\newcommand{\overlr}{\overleftrightarrow}
\newcommand{\AdrHEPC}{Phenikaa Institute for Advanced Study and Faculty of Basic Science, Phenikaa University, Yen Nghia, Ha Dong, Hanoi 100000, Vietnam}

\title{Dark matter in the fully flipped 3-3-1-1 model}

\author{Duong Van Loi}
\email{loi.duongvan@phenikaa-uni.edu.vn} 
\author{Cao H. Nam}
\email{nam.caohoang@phenikaa-uni.edu.vn}
\author{Phung Van Dong} 
\email{Corresponding author; dong.phungvan@phenikaa-uni.edu.vn}
\affiliation{\AdrHEPC}
 
\date{\today}

\begin{abstract}
We present the features of the fully flipped 3-3-1-1 model and show that this model leads to dark matter candidates naturally. We study two dark matter scenarios corresponding to the triplet fermion and singlet scalar candidates, and we determine the viable parameter regimes constrained from the observed relic density and direct detection experiments.
\end{abstract}


\maketitle

\section{Introduction}
The extension of the Standard Model (SM) gauge symmetry based upon the higher weak-isospin symmetry $SU(3)_L$ is best known for solving the number of observed fermion generations, generally called 3-3-1 model when including color and electric charges \cite{Pisano:1991ee,Frampton:1992wt,Singer:1980sw,Montero:1992jk}. In order to cancel the $[SU(3)_L]^3$ anomaly \cite{Gross:1972pv,Georgi:1972bb,Banks:1976yg,Okubo:1977sc}, the traditional arrangement is that one of three quark generations transforms under $SU(3)_L$ differently from remaining two quark generations, whereas all lepton generations transform identically under this group. However, recently Fonseca and Hirsch have made an interesting proposal where one of three lepton generations to transform differently from remaining two lepton generations, while all quark generations are identical under $SU(3)_L$ \cite{Fonseca:2016tbn}. In this way, the 3-3-1 matter content is flipped (i.e. reversed), called the flipped 3-3-1 model. The key for flipping is that the $[SU(3)_L]^3$ anomaly induced by a fermion sextet equals that induced by seven fermion triplets, where note that the color number is not counted. Furthermore, the authors have shown that the flipped 3-3-1 model only realizes a unique and minimal setup of fermion content (see also \cite{Huong:2019vej}). This flip converts the flavor matters in quark sector to the lepton sector as well as leads to the anomalies in interaction of neutrinos with matter \cite{Huong:2019vej}. The lepton flavor violating decays of charged leptons and the SM-like Higgs boson in this model were investigated, all of which were well below experimental bounds \cite{Fonseca:2016tbn,Huong:2019vej,Hong:2020qxc}.

It is noteworthy that the electric charge $Q$ and the baryon-minus-lepton charge $B-L$ neither commute nor close algebraically with $SU(3)_L$ \cite{Huong:2019vej}. To close the symmetries on which we can base our theory, the isospin symmetry $SU(3)_L$ must be enlarged (i.e. flipped) to $SU(3)_L \otimes U(1)_X \otimes U(1)_N$ by symmetry principles, called 3-3-1-1 when including the color group \cite{Dong:2013wca,Dong:2015yra}. Here the new abelian charges $X$ and $N$ are related to $Q$ and $B-L$ through the Cartan generators of $SU(3)_L$, respectively. Hence, the fully flipping in gauge sector yields a complete gauge symmetry and it may of course contain the above flipped fermion content. In this work, we discuss a model that is based on this complete gauge symmetry with the flipped fermion content, called fully flipped 3-3-1-1 model. We show that the model supplies the novel schemes of single-component dark matter (DM). 

The rest of this paper is organized as follows. In Sec. \ref{model}, we provide the features of the model. The mass spectra for the scalar and gauge boson sectors are considered in Sec. \ref{scalarandgauge}. In Sec. \ref{interaction} we compute relevant interactions. Sec. \ref{dark} is devoted to the DM observables. Finally, we conclude this work in Sec. \ref{conl}.

\section{\label{model}The fully flipped 3-3-1-1 model}

As stated, the 3-3-1-1 gauge symmetry is given by
\be SU(3)_C\otimes SU(3)_L\otimes U(1)_X\otimes U(1)_N,\ee
where $SU(3)_C$ is the ordinary color group. The new charges $X$ and $N$ respectively determine $Q$ and $B-L$ as related to the diagonal generators of the $SU(3)_L$ group to be \cite{Huong:2019vej} 
\be Q = T_3 + \frac{1}{\sqrt3} T_8 + X,\hs B-L = \frac{2}{\sqrt3} T_8 + N. \ee

The flipped fermion content under the gauge symmetry $SU(3)_C\otimes SU(3)_L\otimes U(1)_X\otimes U(1)_N$ is arranged, such as
\bea
\psi_{1L} &=& \left(  \begin{array}{ccc} 
      \xi^{+} & \fr{1}{\sqrt{2}}\xi^{0} & \fr{1}{\sqrt{2}} \nu_1 \\
      \fr{1}{\sqrt{2}} \xi^{0} & \xi^{-} & \fr{1}{\sqrt{2}} e_1 \\
      \fr{1}{\sqrt{2}} \nu_1 & \fr{1}{\sqrt{2}} e_1 & E_1 \\
   \end{array}\right)_L  \sim (1,6,-1/3,-2/3),\\ 
\psi_{\alpha L} &=& (\nu_\al\ e_\al\ E_\al)^T_{L} \sim (1,3,-2/3,-4/3),\\ 
Q_{aL} &=& (d_a\ -u_a\ U_a)^T_{L} \sim (3,3^*,1/3,2/3),\\
\nu_{aR} &\sim & (1,1,0,-1),\hs e_{aR} \sim (1,1,-1,-1),\hs E_{aR} \sim (1,1,-1,-2),\\
u_{aR} &\sim & (3,1,2/3,1/3),\hs d_{aR} \sim (3,1,-1/3,1/3),\hs U_{aR} \sim (3,1,2/3,4/3).
\eea Here $\al=2,3$ and $a=1,2,3$ are generation indices, the component fields $\xi$'s compose a real triplet of $SU(2)_L$, and the fields $E,U$ have the same electric charge as $e,u$, respectively.
The above fermion content is free from all the anomalies, including the gravitational anomaly. 

To break the gauge symmetry and generate appropriate masses for the particles, the scalar content is introduced as
\bea \eta &=&(\eta^0_1\ \eta^-_2\ \eta^-_3)^T\sim (1,3,- 2/3,- 1/3),\\ 
\rho &=& (\rho^+_1\ \rho^0_2\ \rho^0_3)^T\sim (1,3, 1/3,- 1/3),\\ 
\chi &=& (\chi^+_1\ \chi^0_2\ \chi^0_3)^T\sim (1,3,1/3,2/3),\\
S &=& \left(\begin{array}{ccc}
S^{++}_{11} & \fr{1}{\sqrt{2}}S^{+}_{12}  & \fr{1}{\sqrt{2}}S^+_{13} \\
\fr{1}{\sqrt{2}}S^+_{12} & S^0_{22} & \fr{1}{\sqrt{2}} S^0_{23}\\
\fr{1}{\sqrt{2}}S^+_{13} & \fr{1}{\sqrt{2}} S^0_{23} & S^0_{33}\\
\end{array}\right)\sim (1,6, 2/3,4/3),\\ 
\phi &\sim& (1,1,0,2),
\eea
which have the corresponding vacuum expectation values
(VEVs) to be
\bea \langle \eta\rangle &=& \fr{1}{\sqrt{2}}\left(	\begin{array}{c}
u\\
0\\
0\\
\end{array}	\right),
\hs  \langle\rho\rangle =\fr{1}{\sqrt{2}}\left(	\begin{array}{c}
0\\
v\\
0\\
\end{array}	\right),
\hs \langle \chi\rangle =\fr{1}{\sqrt{2}}\left(	\begin{array}{c}
0\\
0\\
w\\
\end{array}	\right),\\ 
\langle S\rangle &=& \fr{1}{\sqrt{2}}\left(\begin{array}{ccc}
0 & 0 & 0\\
0 & \kappa & 0\\
0 & 0 & \La \\
\end{array}\right),\hs
\langle\phi\rangle = \fr{1}{\sqrt{2}}\Delta,\eea
where we assume $\kappa\ll u,v\ll w,\Lambda\ll\Delta$. Notice that the other neutral scalars are $W_P$-odd, hence possessing vanished VEV due to the $W_P$ conservation, as shown below.    

The gauge symmetry is broken via three steps, 
\bc
\begin{tabular}{c}
$SU(3)_C\otimes SU(3)_L\otimes U(1)_X \otimes U(1)_N$\\ 
$\downarrow \Delta $\\ 
$SU(3)_C\otimes SU(3)_L \otimes U(1)_X\otimes W'_P$\\ 
$\downarrow w,\Lambda$\\ 
$SU(3)_C\otimes SU(2)_L \otimes U(1)_Y\otimes W_P$\\ 
$\downarrow u,v,\kappa$\\ 
$SU(3)_C\otimes U(1)_Q\otimes W_P$ 
\end{tabular}
\ec
Here $W'_P$ is a residual gauge symmetry of $U(1)_N$ which is continuously broken along with $SU(3)_L$ defining a final residual symmetry $W_P=(-1)^{3(B-L)}$ with $B-L=(2/\sqrt{3})T_8+N$ \cite{Dong:2013wca,Dong:2015yra}. When including the spin symmetry $(-1)^{2s}$, we have \be W_P=(-1)^{3(B-L)+2s}.\ee This symmetry divides the model particles into two classes, as displayed in Table \ref{Wcharge}, where the new fermions and some other particles are odd under $W_P$, while all the remaining particles, including the SM ones, are even under this symmetry. Since $W_P$ is conserved, the odd particles are only coupled in pairs in interactions and the lightest odd particle is stabilized and thus can be a DM candidate if it is electrically neutral and colorless.
\begin{table}
\centering
\begin{tabular}{lccccccccccccc}
\hline\hline
Particle & $\nu$ & $e$ & $u$ & $d$ & $W$ & $\gamma$ & $Z,Z',Z''$ & $\eta_{1,2}$ & $\rho_{1,2}$ & $\chi_3$ & $S_{11,12,22}$& $S_{33}$& $\phi$\\
\hline  
$B-L$ & $-1$ & $-1$ & $1/3$ & $1/3$ & $0$ & $0$ & $0$ & $0$ &$0$& $0$ & $2$ & $0$& $2$\\
$W_P$ & $1$ & $1$ & $1$ & $1$ & $1$ & $1$ & $1$ & $1$ & $1$ & $1$ & $1$ & $1$ & $1$\\
\hline\hline
Particle & $\xi^{+}$ & $\xi^{0}$ & $\xi^{-}$ & $E^-$ & $U^{2/3}$ & $X^+$ & $Y^0$ & $\eta^-_3$ & $\rho^0_3$ & $\chi^+_{1}$ & $\chi^0_2$ & $S^+_{13}$ & $S^0_{23}$\\
\hline  
$B-L$ & 0 & 0 & 0 & $-2$ & $4/3$ & $1$ & $1$ & $-1$ & $-1$ & $1$ & 1 & 1 & 1
\\
$W_P$ & $-1$ & $-1$ & $-1$ & $-1$ & $-1$ & $-1$ & $-1$ & $-1$ & $-1$ & $-1$ & $-1$ & $-1$ & $-1$\\
\hline\hline
\end{tabular}
\caption{\label{Wcharge} The residual gauge symmetry $W_P$ separates the model particles into two classes: even particles according to $W_P=1$ and odd particles according to $W_P=-1$.}
\end{table} 

The total Lagrangian consists of \be \mathcal{L}=\mathcal{L}_{\mathrm{kinetic}}+\mathcal{L}_{\mathrm{Yukawa}}-V.\label{adtlt}\ee The first term composes kinetic terms and gauge interactions,  
\bea
\mathcal{L}_{\mathrm{kinetic}} &=& \sum_F i\bar{F}\gamma^\mu D_\mu F + \sum_\Phi (D^\mu \Phi)^\dag (D_\mu \Phi)\crn
&& - \fr 1 4 G_{i\mu\nu}G_i^{\mu\nu}- \fr 1 4 A_{i\mu\nu}A_i^{\mu\nu}- \fr 1 4 B_{\mu\nu}B^{\mu\nu}- \fr 1 4 C_{\mu\nu}C^{\mu\nu},
\eea
where $F$ and $\Phi$ run over all the fermion and scalar multiplets, respectively. Additionally, the covariant derivative and the field strength tensors are determined by
\bea
D_\mu &=& \partial_\mu + ig_s t_i G_{i\mu} + ig T_i A_{i\mu} + ig_X X B_\mu + ig_N N C_\mu,\\
G_{i\mu\nu} &=& \partial_\mu G_{i\nu} - \partial_\nu G_{i\mu} - g_s f_{ijk} G_{j\mu}G_{k\nu},\\
A_{i\mu\nu} &=& \partial_\mu A_{i\nu} - \partial_\nu A_{i\mu} - g f_{ijk} A_{j\mu}A_{k\nu},\\
B_{\mu\nu} &=& \partial_\mu B_\nu - \partial_\nu B_\mu,\hs C_{\mu\nu} = \partial_\mu C_\nu - \partial_\nu C_\mu,
\eea
where we denote $(g_s,g,g_X,g_N)$, $(t_i,T_i,X,N)$, and $(G_i,A_i,B,C)$ to be the coupling constants, the generators, and the gauge bosons corresponding to the 3-3-1-1 subgroups, respectively, and $f_{ijk}$ are $SU(3)$ structure constants.   
 
The second term of (\ref{adtlt}) contains Yukawa interactions, 
\bea
\mathcal{L}_{\mathrm{Yukawa}} &=& h^e_{\alpha a} \bar{\psi}_{\alpha L}\rho e_{aR} + h^E_{\alpha a} \bar{\psi}_{\alpha L}\chi E_{aR} + h^E_{1a} \bar{\psi}_{1L}SE_{aR} + h^\xi \bar{\psi}^c_{1L}\psi_{1L}S \crn
&& + h^u_{ab} \bar{Q}_{aL}\rho^* u_{bR} + h^d_{ab} \bar{Q}_{aL}\eta^* d_{bR} + h^U_{ab} \bar{Q}_{aL}\chi^* U_{bR}\crn
&& + h^\nu_{\alpha b} \bar{\psi}_{\alpha L}\eta\nu_{bR} + h^R_{ab}\bar{\nu}^c_{aR}\nu_{bR}\phi + H.c., \label{addttba}
\eea which were previously presented in \cite{Huong:2019vej}. There, all the fermion masses were properly produced, so we do not refer to them hereafter. 

The last term of (\ref{adtlt}) is the scalar potential, taking the form 
\be V =V (\eta,\rho,\chi,S)+V(\phi,\mathrm{mix}), \ee
where	
\bea V(\eta,\rho,\chi,S)&=&\mu_1^2 \eta^\dag \eta+\mu_2^2	\rho^\dag \rho+\mu_3^2 \chi^\dag \chi+\mu_4^2 \Tr(S^\dag S)\crn
&&+\la_1(\eta^\dag \eta)^2+\la_2(\rho^\dag \rho)^2+\la_3 (\chi^\dag \chi)^2+\la_4\Tr^2(S^\dag S)+\la_5\Tr(S^\dag S)^2\crn
&&+\la_6(\eta^\dag\eta)(\rho^\dag \rho)+(\la_7\eta^\dag \eta+\la_8\rho^\dag \rho)(\chi^\dag\chi)\crn
&&+(\la_9\eta^\dag \eta+\la_{10}\rho^\dag \rho+\la_{11}\chi^\dag \chi)\Tr(S^\dag S)\crn
&&+\la_{12}(\eta^\dag \rho)(\rho^\dag \eta)+\la_{13}(\eta^\dag \chi)(\chi^\dag\eta)+\la_{14}(\chi^\dag\rho)(\rho^\dag \chi)\crn
&&+\la_{15}(\chi^\dag S)(S^\dag \chi)+\la_{16}(\eta^\dag S)(S^\dag \eta)+\la_{17}(\rho^\dag S)(S^\dag \rho)\crn
&&+[\lambda_{18} \eta\rho (S\chi^*)+f_1 \chi^T S^\dag \chi+f_2 \eta \rho \chi + \mathrm{H.c}.],\label{scalar}\\
V(\phi,\mathrm{mix})&=&\mu_5^2 \phi^\dag \phi+\la_{19}(\phi^\dag\phi)^2+[\la_{20}\eta^\dag\eta+\la_{21}\rho^\dag\rho+\la_{22}\chi^\dag\chi\crn
&& +\la_{23}\Tr(S^\dag S)](\phi^\dag \phi)+(\la_{24}\rho^T S^\dag\rho\phi + \mathrm{H.c}.),
\eea
where the parameters $\mu$'s and $f_{1,2}$ have mass dimension, while the couplings $\lambda$'s are dimensionless. Furthermore, the necessary conditions for the scalar potential bounded from below and producing the gauge symmetry breaking are $\mu^2_{1,2,3,4,5}<0$, $\lambda_{1,2,3,19}>0$, and $\la_4+\la_5 >0$.

\section{\label{scalarandgauge}Scalar and gauge sectors}

Because of the condition $\Delta\gg w, V$, the field $\phi$ can be integrated out from the low-energy effective potential of $\eta,\rho,\chi$, and $S$. As a result, the scalar potential below $\Delta$ has a form similar to $V(\eta,\rho,\chi,S)$ where its couplings become effective due to the modification of heavy particles. Indeed, let us expand $\phi=\frac{1}{\sqrt2}(\Delta+H_C+iG_{C})$, where $H_C$ and $G_{C}$ are the new Higgs and Goldstone bosons associate to $U(1)_N$, respectively. The masses of the new Higgs $H_C$ and the new gauge $C$ are proportional to $\Delta$, $m_{H_C}\simeq \sqrt{2\la_{19}}\Delta$ and $m_{C}\simeq 2 g_N \Delta$, which are decoupled from the low energy particle spectra. Hence, we will neglect the $U(1)_N$ sector.  

Now, we consider the scalar potential (\ref{scalar}). To obtain the potential minimum and physical scalar spectrum, we expand the neutral scalar fields around the VEVs as
\bea \eta &=&\left(	\begin{array}{c}
\fr{1}{\sqrt2}(u+S_1+iA_1)\\
\eta^-_2\\
\eta^-_3\\
\end{array}	\right),\hs \rho =\left(\begin{array}{c}
\rho^+_1\\
\fr{1}{\sqrt2}(v+S_2+iA_2)\\
\fr{1}{\sqrt2}(S_3^\prime+iA_3^\prime)\\
\end{array}\right),\\
 \chi &=&\left(\begin{array}{c}
\chi^+_1\\
\fr{1}{\sqrt2}(S_2^\prime+iA_2^\prime)\\
\fr{1}{\sqrt2}(w+S_3+iA_3)\\	
\end{array}	\right),\hs S=\left(\begin{array}{ccc}
S^{++}_{11} & \fr{1}{\sqrt{2}}S^{+}_{12}  & \fr{1}{\sqrt{2}}S^+_{13} \\
\fr{1}{\sqrt{2}}S^+_{12} &  \fr{1}{\sqrt2}(S_4+iA_4) &\fr{1}{2}(S_1^\prime+iA_1^\prime)\\
\fr{1}{\sqrt{2}}S^+_{13} &\fr{1}{2}(S_1^\prime+iA_1^\prime) & \fr{1}{\sqrt2}(\La +S_5+iA_5) \\	
\end{array}\right),
\eea
where the fields $\rho^0_3$, $\chi^0_2$, and $S^0_{23}$ are odd under $W_P$, cannot develop VEVs, as mentioned. Additionally, since $\kappa$ constrained by the $\rho$-parameter is tiny, its contribution would be neglected. 

Expanding all the terms of the potential (\ref{scalar}) up to the second-order contributions of physical scalar fields, the scalar potential is resulted as $V(\eta,\rho,\chi,S)=V_{\mathrm{min}}+V_{\mathrm{linear}}+V_{\mathrm{mass}}+V_{\mathrm{interaction}}$, where the first term is the potential minimum independent of fields,  
\bea
V_{\mathrm{min}}&=&\fr{1}{4}[2(\mu_1^2 u^2+\mu_2^2 v^2+\mu_3^2 w^2+\mu_4^2 \Lambda^2)+\lambda_1 u^4+\lambda_2 v^4+\lambda_3 w^4+(\lambda_4+\lambda_5)\Lambda^4\crn
&&+\lambda_{6} u^2v^2+\lambda_{7} w^2u^2+\lambda_{8} w^2v^2+(\lambda_9 u^2+\lambda_{10}v^2+\lambda_{11}w^2)\Lambda^2+\lambda_{15} w^2\Lambda^2\crn
&&+2\bar{\La} wuv+2\sqrt2 f_1 w^2\Lambda],
\eea
where $\bar{\La}=\sqrt2 f_2+\lambda_{18}\Lambda$.
The second term $V_{\mathrm{linear}}$ contains all the linear terms in fields, should be vanished due to the gauge invariance, leading to 
\bea
&&[2(\mu_1^2+\lambda_1 u^2)+\lambda_6 v^2 +\lambda_7 w^2 +\lambda_9 \Lambda^2]u+\bar{\La} wv =0,\\
&&[2(\mu_2^2+\lambda_2 v^2)+\lambda_6 u^2 +\lambda_8 w^2 +\lambda_{10} \Lambda^2]v+\bar{\La}wu =0,\\
&&[2(\mu_3^2+\lambda_3 w^2)+\lambda_7 u^2 +\lambda_8 v^2 +(\lambda_{11}+\lambda_{15}) \Lambda^2]w+\bar{\La}uv+2\sqrt2 f_1w \Lambda=0,\\
&&[2(\mu_4^2+\lambda_4\La^2+\lambda_5\Lambda^2)+\lambda_9 u^2 +\lambda_{10} v^2 +(\lambda_{11}+\lambda_{15}) w^2]\La+\Om w=0,
\eea
where $\Om=\sqrt2 f_1 w+\lambda_{18} uv$. These minimization conditions yield the VEVs $u,v,w,\La$ as desirable. 

The mass part $V_{\mathrm{mass}}$ consists of the quadratic terms in fields, grouped into
$V_{\mathrm{mass}}=V_{\mathrm{mass}}^{S}+V_{\mathrm{mass}}^{A}+V_{\mathrm{mass}}^{\mathrm{charged}}+V_{\mathrm{mass}}^{S^\prime}+V_{\mathrm{mass}}^{A^\prime}$.
Here the charged term includes charged scalars, while the remaining terms describe $CP$-even and $CP$-odd scalar fields, and notice that the primed fields are decoupled from the normal fields due to the $W_P$ conservation. That said, after integrating $\phi$ out, the scalar potential (\ref{scalar}) gives a mass spectrum of the scalar sector including 22 massive Higgs fields, summarized as $H^0_{1,2,3,4}$, $H^{0,0*}_5$, $\mathcal{A}^0_{1,2}$, $H_{11}^{\pm\pm}$, $H_{12}^\pm$, $H^\pm_1$, $\mathcal{H}'^{\pm}_{1,2}$, and $H'^{0,0*}_{1,2}$, where $H_1^0$ is identified as the SM-like Higgs boson with the mass in the weak scale, while the others are new Higgs bosons with the masses at $w,\Lambda$ scales. Additionally, there are 8 Goldstone bosons, determined as $G_Z$, $G_{Z'}$, $G_W^\pm$, $G_X^\pm$, and $G_Y^{0,0*}$, correspondingly eaten by 8 massive gauge bosons. 

The mentioned Higgs and Goldstone bosons are related to those in the gauge basis, such as
\bea
\left(   \begin{array}{c}     S_1 \\     S_2 \\   \end{array} \right) &=&  \left(   \begin{array}{cc}   c_{\al_1} & s_{\al_1} \\  s_{\al_1} & -c_{\al_1} \\   \end{array}\right) \left(   \begin{array}{c}    H_1 \\    H_2 \\  \end{array}\right),\hs \left(   \begin{array}{c}    S_3 \\    S_5 \\  \end{array}\right) = \left(  \begin{array}{cc}  c_{\al_2} & s_{\al_2} \\  -s_{\al_2} & c_{\al_2}\\  \end{array} \right)\left(  \begin{array}{c}    H_3 \\    H_4 \\  \end{array}\right),\crn
\left(  \begin{array}{c}  A_1 \\ A_2 \\ A_3 \\  A_5 \\  \end{array}\right) &\simeq & \left(\begin{array}{cccc} c_{\al_1} & \frac{s_{\alpha_1}c_{\alpha_1}v}{\sqrt{w^2+4\La^2}} & -\sqrt{\frac{w^2+4\La^2}{u^2+v^2}}s_{\al_1} s_{\ep_1} & \sqrt{\frac{w^2+4\La^2}{u^2+v^2}}s_{\al_1} c_{\ep_1}   \\  
 -s_{\al_1} &   \frac{s_{\alpha_1}c_{\alpha_1}u}{\sqrt{w^2+4\La^2}}  & -\sqrt{\frac{w^2+4\La^2}{u^2+v^2}}c_{\al_1}s_{\ep_1} & \sqrt{\frac{w^2+4\La^2}{u^2+v^2}}c_{\al_1} c_{\ep_1} \\
0 & -s_{\al_3} & c_{\ep_1}c_{\al_3}-s_{\ep_1}s_{\al_3}s_{\al_1}c_{\al_1} &  c_{\ep_1}s_{\al_3}s_{\al_1}c_{\al_1}+s_{\ep_1}c_{\al_3}\\ 
0& -c_{\al_3} & -c_{\ep_1}s_{\al_3}-s_{\ep_1}c_{\al_3}s_{\al_1}c_{\al_1} & c_{\ep_1}c_{\al_3}s_{\al_1}c_{\al_1}-s_{\ep_1}s_{\al_3} \\  \end{array}\right)\left(  \begin{array}{c}  G_Z \\ G_{Z'} \\ \mathcal{A}_1 \\  \mathcal{A}_2 \\  \end{array}\right), \crn
\left(  \begin{array}{c}  \eta_2^\pm \\  \rho_1^\pm \\ \end{array}\right) &\simeq & \left(\begin{array}{cc}c_{\al_1} & s_{\al_1} \\  -s_{\al_1} &  c_{\al_1}\\  \end{array}\right)\left(  \begin{array}{c}     G_W^\pm \\   H_1^\pm  \\  \end{array}\right), \label{rscalar} \\
\left(  \begin{array}{c}  \eta_3^\pm \\  \chi_1^\pm \\ S_{13}^\pm \\ \end{array}\right) &\simeq & \left(\begin{array}{ccc} \fr{u}{\sqrt{w^2+2\La^2}} & c_{\ep_2} & s_{\ep_2}  \\ 
-s_{\al_4}& \fr{s_{\al_4}c_{\ep_2}u}{\sqrt{w^2+2\La^2}}- c_{\al_4}s_{\ep_2} & c_{\al_4}c_{\ep_2}  \\  
-c_{\al_4} & \fr{c_{\al_4}c_{\ep_2}u}{\sqrt{w^2+2\La^2}}+ s_{\al_4}s_{\ep_2}  & - s_{\al_4}c_{\ep_2}  \\  \end{array}\right)\left(  \begin{array}{c}     G_X^\pm \\   \mathcal{H}'^{\pm}_1  \\  \mathcal{H}'^{\pm}_2 \\  \end{array}\right), \crn
\left(  \begin{array}{c}  \chi_2^0 \\  \rho_3^{0*} \\ S_{23}^0 \\ \end{array}\right) &\simeq & \left(\begin{array}{ccc} s_{\al_4} & c_{\al_4}s_{\ep_3}+\frac{v}{w}c_{\ep_3} & -c_{\al_4} c_{\ep_3} \\-\frac{v}{\sqrt{w^2+2\La^2}} & c_{\ep_3} & s_{\ep_3}+\frac{v}{w}c_{\al_4}c_{\ep_3}\\ c_{\al_4} & -s_{\al_4}s_{\ep_3} & s_{\al_4} c_{\ep_3} \\  \end{array}\right)\left(  \begin{array}{c}  G_Y^0 \\ H'^{0}_1  \\ H'^{0}_2  \\  \end{array}\right),\crn
S_{22}^0&\equiv & H_5^0,\hs S_{12}^\pm \equiv H_{12}^\pm, \hs S_{11}^{\pm\pm} \equiv H_{11}^{\pm\pm},\nn 
\eea
where we define $s_{\al_1}\equiv\sin \al_1$, $c_{\al_1}\equiv\cos \al_1$, $\tan \al_1\equiv t_{\al_1}=v/u$, and
\bea
t_{\al_2} &=& \frac{4[(\la_{11}+\la_{15})\La+\sqrt2 f_1]w\La}{4(\la_4+\la_5)\La^3-(4\la_3\La+\sqrt2 f_1)w^2},\hs t_{\al_3} = \frac{w}{2\Lambda},\hs t_{\al_4} = \frac{w}{\sqrt2\Lambda},\crn 
t_{2\ep_1}&\simeq &\frac{2[\la_{18}(w^2+2\La^2)-2\sqrt2 f_2\La]\La uv\sqrt{u^2+v^2}}{[\sqrt2 f_1(w^2+4\La^2)uv-\bar{\La}w\La (u^2+v^2)]\sqrt{w^2+4\La^2}},\crn
t_{2\ep_2}&\simeq &\frac{2\sqrt2[(2\la_{13}-\la_{16})w\La u+(\la_{18}w^2-2\sqrt2 f_2\La)v]\La u}{\sqrt{w^2+2\La^2}\{[(2\la_{13}+\la_{15})w^2+2(\la_{15}+\la_{16})\La^2]\La u-2\bar{\La}w\La v+2\sqrt2 f_1(w^2+2\La^2)u\}},\crn
t_{2\ep_3}&\simeq &\frac{-2\sqrt2[(\la_{15}+\la_{17})\La v-\la_{18}wu+2\sqrt2 f_1v]\La v \sqrt{w^2+2\La^2}}{w[(2\la_{14}+\la_{15})w^2\La v+2(\la_{15}+\la_{17})\La^3v-2\bar{\La}w\La u+2\sqrt2 f_1(w^2+2\La^2)v]}.
\eea

The masses of the mentioned Higgs fields (the masses of the mentioned Goldstone bosons vanish, thus unlisted) are given by
\bea
m^2_{H_1}&\simeq &\frac{2}{u^2+v^2}(\la_1u^4+\la_2v^4+\la_6u^2v^2),\hs m^2_{H_2}\simeq -\frac{1}{s_{2\al_1}}(\sqrt2 f_2+\lambda_{18}\Lambda)w,\\
m^2_{H_{3,4}}&\simeq& \left(\la_3-\frac{f_1}{2\sqrt2 \La}\right)w^2+(\la_4+\la_5)\La^2\crn
&&\mp\sqrt{[(\la_{11}+\la_{15})\La+\sqrt2 f_1]^2w^2+\left[(\la_4+\la_5)\La^2-\left(\la_3+\fr{f_1}{2\sqrt2 \La}\right)w^2\right]^2},\\
m^2_{H_5}&=&-\fr{1}{2\La}\left[(2\la_5\La^2+\la_{15}w^2-\la_{17}v^2)\La+(\sqrt2 f_1 w +\la_{18}uv) w\right],\\
m^2_{\mathcal{A}_1}&\simeq & -\frac{2\sqrt2}{c^2_{\al_3}} f_1\La,\hs m^2_{\mathcal{A}_2}\simeq -\frac{1}{s_{2\al_1}}(\sqrt2 f_2 + \la_{18}\Lambda)w,\\
m^2_{H_{11}^{\pm\pm}}&=&-\fr{1}{2\La}\left[(2\la_5\La^2+\la_{15}w^2-\la_{16}u^2)\La+(\sqrt2 f_1 w +\la_{18}uv) w\right],\\
m^2_{H_{12}^{\pm}}&=&-\fr{1}{4\La}\left[(4\la_5\La^2+2\la_{15}w^2-\la_{16}u^2-\la_{17}v^2)\La+2(\sqrt2 f_1 w +\la_{18}uv) w\right],\\
 m^2_{H_1^\pm}&=&\frac{1}{s_{2\al_1}}\left[\la_{12}uv-(\sqrt2 f_2 + \la_{18}\Lambda)w\right],\\
 m^2_{\mathcal{H}'^{\pm}_1}&\simeq &\frac{1}{2}\left[\la_{13}w^2+\la_{16}\La^2-(\sqrt2 f_2 + \la_{18}\Lambda)w t_{\al_1}\right],\\
 m^2_{\mathcal{H}'^{\pm}_2}&\simeq &-\fr{1}{2c^2_{\al_4}}(\la_{15}\La+2\sqrt2 f_1)\La,\\
 m_{H'_1}^2&\simeq & \fr{1}{2}\left[\la_{14}w^2+\la_{17}\La^2-\frac{u}{v}(\sqrt2 f_2 + \la_{18}\Lambda)w\right],\\
 m_{H'_2}^2&\simeq & -\fr{1}{2c^2_{\al_4}}(\la_{15}\La+2\sqrt2 f_1)\La.
\eea

For the gauge boson sector, recall that the $U(1)_N$ gauge boson is heavy, which is integrated out as $\phi$ is. Hence, the spectrum of the remaining gauge bosons is identical to those obtained in \cite{Huong:2019vej}. That said, we have two new non-Hermitian gauge bosons $X,Y$ with the masses at the $w, \Lambda$ scales, besides the $W$ boson of the SM, as follows
\bea
W^\pm_\mu &=& \frac{1}{\sqrt2} (A_{1\mu}\mp i A_{2\mu}), \hs X^\pm_\mu = \frac{1}{\sqrt2} (A_{4\mu}\mp i A_{5\mu}), \hs Y^{0,0*}_\mu = \frac{1}{\sqrt2} (A_{6\mu}\mp i A_{7\mu}),\label{rcgauge}\\
m^2_W &\simeq& \frac{g^2}{4}(u^2+v^2), \hs m^2_X = \frac{g^2}{4}(u^2+w^2+2\Lambda^2), \hs m^2_Y \simeq \frac{g^2}{4}(v^2+w^2+2\Lambda^2). \label{tvld12}
\eea
The mass of $W$ implies $u^2+v^2\simeq (246\ \mathrm{GeV})^2$. For the neutral gauge sector, the physical fields are related to the gauge fields as $(A_{3\mu} \ A_{8\mu} \ B_{\mu})^T = U (A_{\mu} \ Z_{\mu} \ Z'_{\mu})^T$, with
\be
U = \left(\begin{array}{ccc} s_W & c_W c_\varphi & c_W s_\varphi \\  
\frac{1}{\sqrt3}s_W  &  - \frac{1}{\sqrt3} s_Wt_Wc_\varphi - \sqrt{1-\fr 1 3 t^2_W}s_\varphi  & -\frac{1}{\sqrt3} s_Wt_Ws_\varphi + \sqrt{1-\fr 1 3 t^2_W}c_\varphi  \\
\sqrt{1-\fr 4 3 s^2_W}  &  \frac{t_W}{\sqrt3}(s_\varphi - \sqrt{3-4s^2_W}c_\varphi)  & -\frac{t_W}{\sqrt3}(c_\varphi + \sqrt{3-4s^2_W}s_\varphi)  \end{array}\right),\label{rngauge}
\ee
where $s_W = \sqrt3 t_X/\sqrt{3+4t^2_X}$, $t_X=g_X/g$, and
\be t_{2\varphi}\simeq \frac{\sqrt{3-t^2_W}}{2c^3_W}\frac{u^2-v^2c_{2W}}{w^2+4\Lambda^2}\ee defines the $Z$-$Z'$ mixing angle, $\varphi$.
Here $A$ is the massless photon field, while $Z$ is the SM neutral weak boson with mass $m_Z\simeq m_W/c_W$. $Z'$ is the new neutral gauge boson, obtaining a mass at $w,\Lambda$ scale, such as \be m^2_{Z'}\simeq \fr{g^2}{3-t^2_W}(w^2+4\La^2).\label{tvld11}\ee

The lower bound of $w,\La$ is given by the rho parameter constraint to be $\sqrt{w^2+4\La^2}\sim 5\mathrm{-}7\ \mathrm{TeV}$ dependent on $u,v$ relation, as well as the LHC search for $Z'$ through dilepton products makes a constraint on $Z'$ mass, $m_{Z'}> 2.25$--2.8 TeV, which translates to \be \sqrt{w^2+4\La^2}=\fr{m_{Z'}}{g}\sqrt{3-t^2_W}> 5.68\mathrm{-}7\ \mathrm{TeV},\label{lddabc}\ee in agreement to the rho parameter \cite{Huong:2019vej}. We made a study deduced from the LEPII bound for the process $e^+e^-\to \mu^+\mu^-$ via $Z'$ exchange \cite{Aaboud:2017buh} which reveals $m_{Z'}> 2.7$ TeV, implying a similar bound for $w,\La$.

\section{\label{interaction} Relevant interactions}
\subsection{Fermion and gauge boson interactions}
The gauge interactions of fermions arise from,
\be \mathcal{L}_f = \sum_F i\bar{F}\gamma^\mu D_\mu F. \ee

Using the result in (\ref{rcgauge}), we get the charged current interactions of $W$, $X$, and $Y$ to be 
\be \mathcal{L}^C_f = J^{-\mu}_W W^+_\mu + J^{-\mu}_X X^+_\mu + J^{0*\mu}_Y Y^0_\mu + H.c., \ee
where the currents are given by
\bea
J ^{-\mu}_W &=& -\frac{1}{\sqrt2}g (\bar{\nu}_{aL}\gamma^\mu e_{aL}+\bar{u}_{aL}\gamma^\mu d_{aL}),\\
J^{-\mu}_X &=& -\frac{1}{\sqrt2}g (\sqrt2 \bar{\nu}_{1L}\gamma^\mu E_{1L}+\bar{\nu}_{\alpha L}\gamma^\mu E_{\alpha L} - \bar{U}_{aL}\gamma^\mu d_{aL} + \sqrt2 \bar{\xi}^+_L\gamma^\mu\nu_{1L} + \bar{\xi}^0_L\gamma^\mu e_{1L}),\\
J^{0*\mu}_Y &=& -\frac{1}{\sqrt2}g (\sqrt2 \bar{e}_{1L}\gamma^\mu E_{1L}+\bar{e}_{\alpha L}\gamma^\mu E_{\alpha L} + \bar{U}_{aL}\gamma^\mu u_{aL} + \bar{\xi}^0_L\gamma^\mu\nu_{1L} + \sqrt2 \bar{\xi}^-_L\gamma^\mu e_{1L}).
\eea

Using the result in (\ref{rngauge}), we get the neutral current interactions for the neutral gauge bosons,
\bea 
\mathcal{L}^N_f &=& -eQ(f)\bar{f}\gamma^\mu f A_\mu - \frac{g}{2c_W}\{\bar{f}\gamma^\mu [g^Z_V(f)-g^Z_A(f)\gamma_5]f Z_\mu +\bar{f}\gamma^\mu [g^{Z'}_V(f)-g^{Z'}_A(f)\gamma_5]f Z'_\mu \},
\eea
where $e=gs_W$ and $f$ refers to every fermion of the model, except for the right-handed neutrinos. The vector and axial-vector couplings of $Z$ are collected in Table \ref{Zgiffermion} as put in Appendix \ref{coupscalars}. Notice that all the interactions between $Z$ boson and ordinary fermions are consistently recovered in the limit $\varphi\to 0$.
 
The couplings of $Z'$ with fermions can be obtained from those for $Z$, by replacing \be g^{Z'}_{V,A}(f) = g^{Z}_{V,A}(f) (c_\varphi\to s_\varphi, s_\varphi\to -c_\varphi),\ee which need not necessarily to be determined, but pointed out in the Table \ref{Zgiffermion} caption. 

Notice that $\nu_R$'s have only gauge interaction with the $U(1)_N$ gauge boson and are obviously integrated out as $C$ is, since they possess a large mass proportional to $\Delta$ via the coupling $\nu_R\nu_R\phi$~\cite{Huong:2019vej}.

\subsection{Scalar and gauge boson interactions}

The gauge interactions of the scalar fields arise from
\be \mathcal{L}_\Phi = \sum_\Phi (D^\mu\Phi)^\dag (D_\mu\Phi). \ee 
Substituting the physical fields from (\ref{rscalar}), (\ref{rcgauge}), and (\ref{rngauge}) to this Lagrangian, we get the interactions between a gauge boson and two scalars, two gauge bosons and a scalar, and two gauge bosons and two scalars in the model, given in \ref{1NG2STab1}, \ref{1CG2STab1}, \ref{1S2GTab1}, \ref{1NG1CG2STab1}, \ref{1NG1CG2STab2}, \ref{1NG1CG2STab3}, \ref{2CG2STab1}, \ref{2CG2STab2}, \ref{2NG2STab1}, and \ref{2NG2STab2}, which are gathered in Appendix \ref{coupscalars}. At the leading order, we have verified that all the interactions between the SM-like Higgs boson and the gauge bosons are consistently recovered. However, only the interactions relevant to the following diagrams are needed, yielding the cross-sections as given.    

\section{\label{dark}Search for DM}

In this model, the two scalars $H^{\prime 0}_{1,2}$, the fermion $\xi^0$, and the gauge boson $Y^0$, which all have masses at $w,\Lambda$ scale, are $W_P$-odd particles and electrically neutral and colorless. Hence, they may be responsible for the DM candidate. As indicated in \cite{Dong:2013wca} and updated in \cite{Nam:2020twn}, the relic density contribution of the gauge boson $Y^0$ is small compared to the observed dark matter density, so we do not interpret the vector field as DM. Additionally, if the scalar $H'^0_2$ that transforms as a $SU(2)_L$ doublet has a correct relic density to be DM, it should be ruled out by the direct detection experiments due to a large scattering cross-section induced by $Z$ \cite{Cirelli:2005uq,Barbieri:2006dq}. Therefore, in the following we study the DM phenomenology associated with the candidates, singlet scalar $H^{\prime 0}_1$ and triplet fermion $\xi^0$. The presence of the triplet candidate, a result of the fully flipped, would make dark matter phenomena of the model completely different from the other theories of this type, such as the ordinary 3-3-1 \cite{Dong:2013ioa,Mizukoshi:2010ky}, 3-3-1-1 \cite{Dong:2014wsa,CarcamoHernandez:2020ehn,Leite:2020bnb}, 3-2-3-1 \cite{Huong:2018ytz}, and flipped trinification \cite{Dong:2017zxo}. 
 
\subsection{DM as a singlet scalar}
\begin{figure}[h]
	\centering
	\includegraphics[scale=0.9]{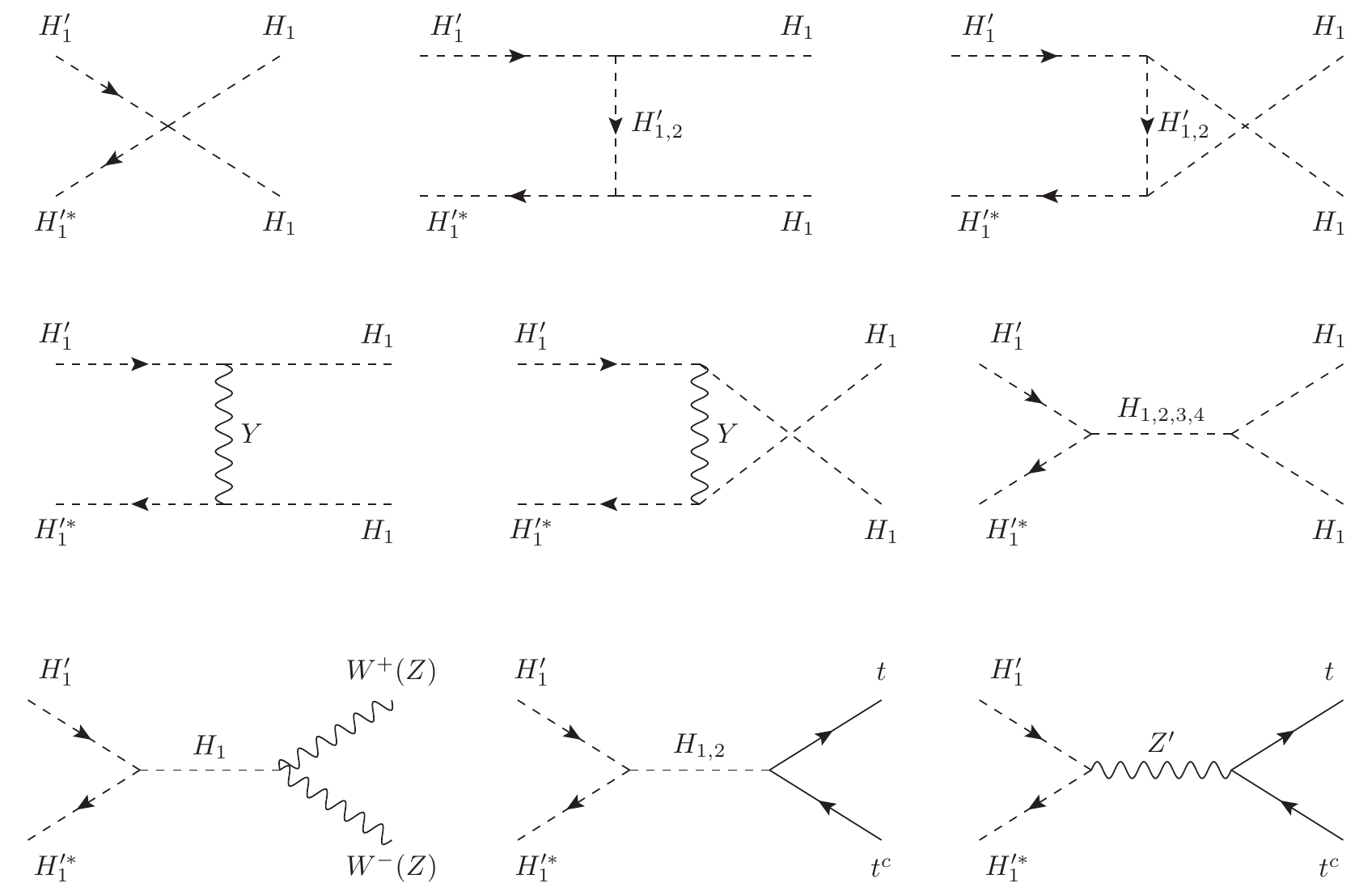}
	\caption{\label{anniH10} Relevant contributions to the scalar DM pair annihilation into SM particles.}	
\end{figure}

We now consider the DM scenario where $H'^0_1$ is the lightest particle among all of the $W_P$-odd particles. The DM pair annihilation into the SM particles, via the most relevant channels, is described in Figure \ref{anniH10}.

Corresponding to each process, the thermal average annihilation cross-section times the relative velocity is approximately given by
\bea
\langle\sigma v\rangle_{H'_1H'_1\to H_1H_1}&=& \frac{1}{16\pi m^2_{H'_1}}\left\{ 2\la_2 s^2_{\alpha_1}+\la_6 c^2_{\alpha_1}\right.\crn
&& + \frac{2[\sqrt2 f_2 c_{\al_1}c_{\al_4}-ws_{\al_4}(0.5\la_{14}c_{\al_4}+\la_{18}c_{\al_1})]^2}{m^2_{H'_1}+m^2_{H'_2}}\crn
&&  +\frac{g^2 s^2_{\al_1}m^2_{H'_1}}{2(m^2_{H'_1}+m^2_Y)}\left(5+\frac{m^2_{H'_1}}{m^2_Y}\right) \crn
&& -3\left[(2\la_1 -\la_6) v c^3_{\al_1} -(2\la_2-\la_6) u s^3_{\al_1}\right] \fr{\la_6 u s_{\al_1}-2\la_2 v c_{\al_1}}{4m^2_{H'_1}-m^2_{H_2}}\crn
&&- \left[(\la_7 c^2_{\al_1}+\la_8 s^2_{\al_1})w c_{\al_2}-(\la_9 c^2_{\al_1}+\la_{10} s^2_{\al_1})\Lambda s_{\al_2}\right.\crn
&&\left. + s_{\al_1}c_{\al_1} (\la_{18}\Lambda c_{\al_2}-\la_{18}w s_{\al_2}+\sqrt2 f_2 c_{\al_2})\right]\crn
&&\times \frac{(\la_8+\la_{14})wc_{\al_2}-(\la_{10}+\la_{17})\Lambda s_{\al_2}}{4 m^2_{H'_1}-m^2_{H_3}}\crn
&&\left.+ (m_{H_3}\leftrightarrow m_{H_4}, c_{\al_2}\leftrightarrow s_{\al_2}, s_{\al_2}\leftrightarrow -c_{\al_2})\right\}^2, \label{tdlt21}\\
\langle\sigma v\rangle_{H'_1H'_1\to W^+W^-}&=& \frac{(2\la_2 v^2+\la_6 u^2)^2}{8\pi (u^2+v^2)^2m^2_{H'_1}},\hs \langle\sigma v\rangle_{H'_1H'_1\to ZZ}= \frac{(2\la_2 v^2+\la_6 u^2)^2}{16\pi (u^2+v^2)^2m^2_{H'_1}},\\
\langle\sigma v\rangle_{H'_1H'_1\to tt^c}&=& \frac{3m_t^2}{\pi m^2_{H'_1}(u^2+v^2)^2}\left[\fr{(2\la_2 v^2+\la_6 u^2)^2}{16m^2_{H'_1}}+\fr{(2\la_2 -\la_6)^2m^2_{H'_1}u^4}{(4m^2_{H'_1}-m^2_{H_2})^2}\right],
\eea
where the annihilation channels into the SM-like Higgs bosons ($H_1$) involving the $H'_1$, $H_1$ propagators as well as the last annihilation channel exchanged by $Z'$ are infinitesimal, as neglected, because of the small coupling or non-relativistic dark matter momentum suppression. Indeed, although the channels exchanged by $H'_{1,2}$ are the same, the $H_1 H'_1 H'^*_1$ coupling as proportional to $u,v$ is radically smaller than the $H_1 H'_1 H'^*_2$ coupling as proportional to $w,f_2$. So the channels exchanged by $H'_1$ are suppressed as compared to those by $H'_2$, for which only the $H'_2$ contribution appears in Eq. (\ref{tdlt21}). Note that the $H_2 H'_1 H'^*_1$ coupling is also proportional to $u,v$. Hence, for the same reason, the $s$-channel contributions of $H_{1,2}$ to the process $H'_1H'_1\rightarrow H_1 H_1$ are much smaller than those by $H_{3,4}$. The SM Higgs ($H_1$) contribution is straightforwardly prevented, while the $H_2$ one may become significant, as kept, due to a resonance in the relic density when the DM mass is close to its mass. The amplitude of the $Z'$-exchanged diagram is proportional to the DM momentum $\vec{p}\simeq m_{H'_1}\vec{v}$ at the $Z' H'_1 H'^*_1$ vertex, hence strongly-suppressed by $|\vec{v}|\sim 10^{-3}$. Additionally, the DM stability requires $m_{H'_1}<m_X\simeq m_Y<m_{Z'}$, where the last two comparisons are derived from Eqs. (\ref{tvld12}) and (\ref{tvld11}) for $w,\La\gg u,v$. The $Z'$ amplitude is also suppressed by $1/m^2_{Z'}$ through its propagation. With the two suppressions, this contribution is manifestly omitted in comparison to the dominant channels.\footnote{There may exist an extremely narrow resonance in the relic density by $Z'$ mass close to the DM stable limit, i.e. $m_{H'_1}=\fr 1 2 m_{Z'}\sim m_{X,Y}$, but it is unreasonable and skipped in this work.} Note that $Z$ does not couple to $H'_1 H'^*_1$ in the effective limit, $(u,v)^2/(w,\La)^2\ll 1$, hence it does not contribute to such channel too. Furthermore, at the effective limit, the $SU(2)_L$ singlets, $H_{3,4}$, do not couple to top quarks. They do not contribute to the annihilation channel to $tt^c$, unlike the $H_{1,2}$ in the above diagram. This $H_2$ correction has a size similar to the $H_2$ contribution in the $H'_1 H'_1\to H_1 H_1$ process, as mentioned. On the other hand, since $H_{2,3,4}$ do not couple to $WW$ and $ZZ$, as seen in Appendix \ref{coupscalars}, they do not contribute to the processes $H'_1 H'_1\rightarrow WW(ZZ)$. There might exist $t$- and/or $u$-channel diagrams associate to the processes $H'_1 H'_1\rightarrow WW(ZZ)$, as exchanged by $X,Y$, but they are all discarded due to the $(u,v)^2/m^2_{X,Y}$ suppression in comparison to the $H_1$ contribution. Also, such corrections are significantly smaller than the $Y$-exchanged diagrams to $H_1 H_1$, which result in Eq. (\ref{tdlt21}). Finally, the relic density of the $H'_1$ is $\Omega_{H'_1}h^2 \simeq 0.1 \mathrm{pb}/\langle\sigma v\rangle$, where 
\be \langle\sigma v\rangle = \langle\sigma v\rangle_{H'_1H'_1\to H_1H_1}+\langle\sigma v\rangle_{H'_1H'_1\to W^+W^-}+\langle\sigma v\rangle_{H'_1H'_1\to ZZ}+\langle\sigma v\rangle_{H'_1H'_1\to tt^c}. \ee

To study the direct detection for the $H'_1$ via the spin-independent (SI) scattering on nuclei, we write the effective Lagrangian describing DM–nucleon interaction in the limit of zero-momentum transfer through the exchange of the Higgs boson $H_1$ as follows
\be \mathcal{L}^\mathrm{eff}_{H'_1-\mathrm{quark}} = \frac{2\sqrt2 m_q}{(u^2+v^2)m^2_{H_1}}(2\la_2 v^2+\la_6 u^2)  H'_1H'_1\bar{q}q. \ee
The SI cross-section for the scattering of $H'_1$ on a target nucleus is given by \cite{Belanger:2008sj}
\be \sigma^\mathrm{SI}_{H'_1N} = \left(\frac{2m_{H'_1N}}{m^2_{H_1}}\frac{m_p}{m_{H'_1}} C_N\right)^2, \label{t1ln321} \ee
where $N=p,n$ and $m_{H'_1N}=m_{H'_1}m_N/(m_{H'_1}+m_N)\simeq m_N$ is the DM-nucleon reduced mass. The nucleus factor, $C_N$, is 
\be 
C_N = \frac{4\sqrt2}{27(u^2+v^2)}(2\la_2 v^2+\la_6 u^2)\left\{3Af^p_{Tg}+\sum_{q=u,d,s}[Zf^p_{Tq}+(A-Z)f^n_{Tq}]\right\},
\ee 
where $Z$ and $A$ correspond to the nucleus charge and the total number of nucleons in the nucleus, respectively, and \cite{Ellis:2000ds}
\bea
f^{p(n)}_{Tu} &\simeq & 0.020 \ (0.014), \hs f^{p(n)}_{Td}\simeq 0.026 \ (0.036),\\
f^{p(n)}_{Ts} &\simeq & 0.118 \ (0.118), \hs f^p_{Tg} = 1-\sum_{q=u,d,s} f^p_{Tq}.
\eea 

For numerical computation in this subsection and the next one for DM fermion, we take the following values of known parameters as \cite{Zyla:2020zbs}
\bea && u = v \simeq 174 \ \mathrm{GeV},\ s^2_W \simeq 0.231,\ \al\simeq 1/128,\ g=\sqrt{4\pi\alpha}/s_W,\crn
&& m_t \simeq 173.1\ \mathrm{GeV},\ m_{H_1} \simeq 125.3 \ \mathrm{GeV},\ A = 131, \ Z=54,\ m_N\simeq 1\ \mathrm{GeV}.\label{pafoscaddttt}\eea

Since the superheavy $U(1)_N$ sector is decoupled, the intermediate new physics regime set by the 3-3-1 breaking scales $w,\La$ is the most relevant to dark matter phenomena. As seen, the scalar dark matter observables are governed by the dark matter mass $m_{H'_1}$ and the new Higgs portals, given through the new Higgs masses and couplings, that connect the dark sector to the normal sector. Additionally, the masses of all such dark matter and new Higgs particles are proportional to the scalar couplings and the scales $w,\La$. Hence, we will investigate two alternative benchmark choices of parameters, either fixing $w,\La$ while changing necessary couplings or varying $w,\La$ while fixing relevant couplings, which might lead to distinct new physics results. 

\subsubsection{Fixing $w,\La$ while varying necessary couplings} 

When the new physics scales $w,\La$ are fixed, the dark matter observables depend on the dimensionless ($\la$'s) and mass-dimension ($f$'s) scalar couplings. Among these couplings, we choose the two typical parameters, $f_2$ and $\la_2$, to be varied, while the remainders are fixed as 
\bea  
\la_6 = -0.2,\ \la_{5,14,17,18} = -0.01,\ \la_{3,4,7,8,9,10,11,12,13,15,16} = 0.5,\ f_1=-w. \label{pafosca}
\eea Note that $\la_1$ is related to $\la_2$ via the SM Higgs mass. Although $\la_2$ can be changed, we will choose only three values of it that characterize its viable range, simultaneously making the DM phenomenology suitable, as supplied in the following figures.    

\mathversion{normal}
\begin{figure}[!h]
\begin{center}
	\includegraphics[scale=0.4]{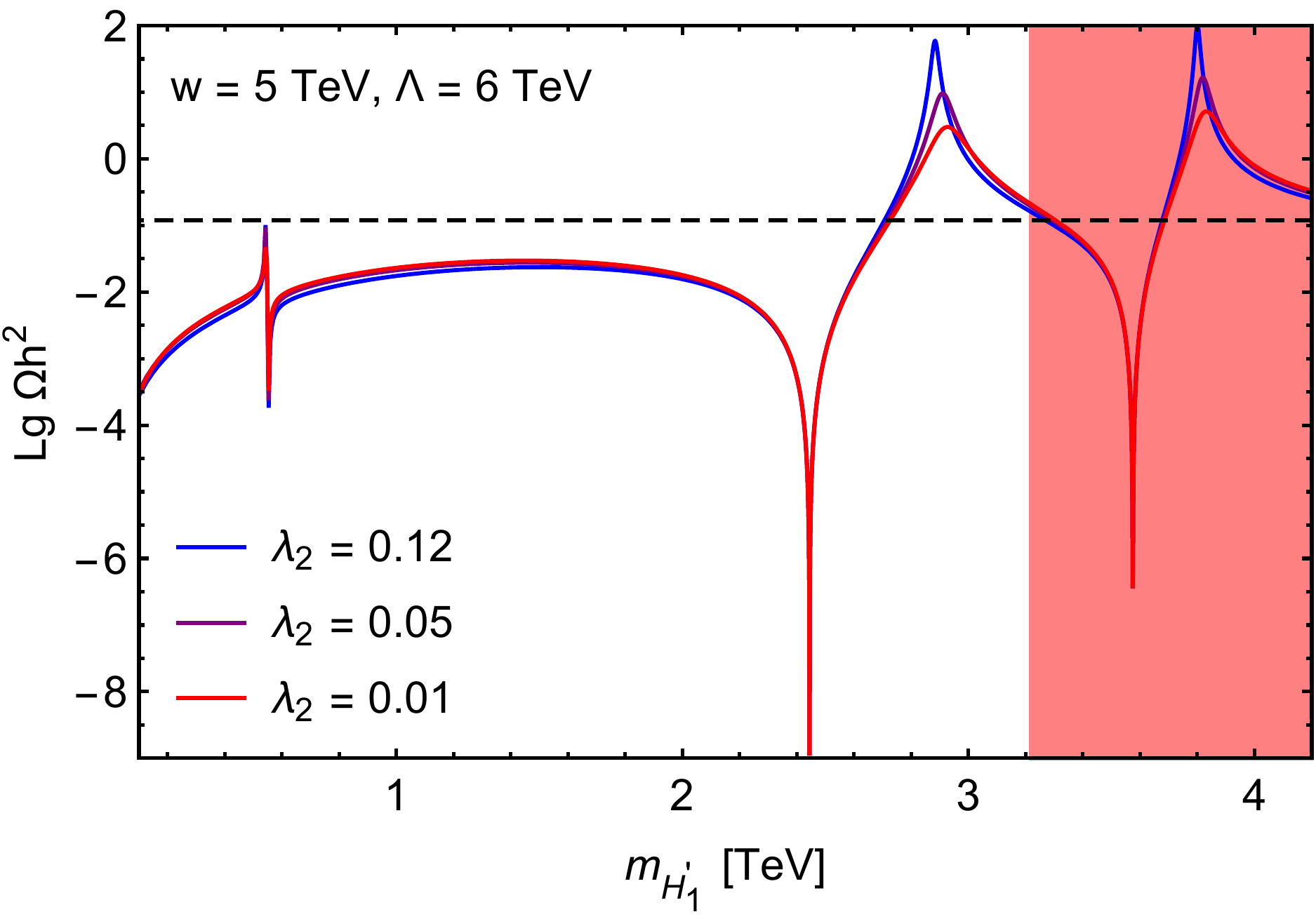}
	\includegraphics[scale=0.4]{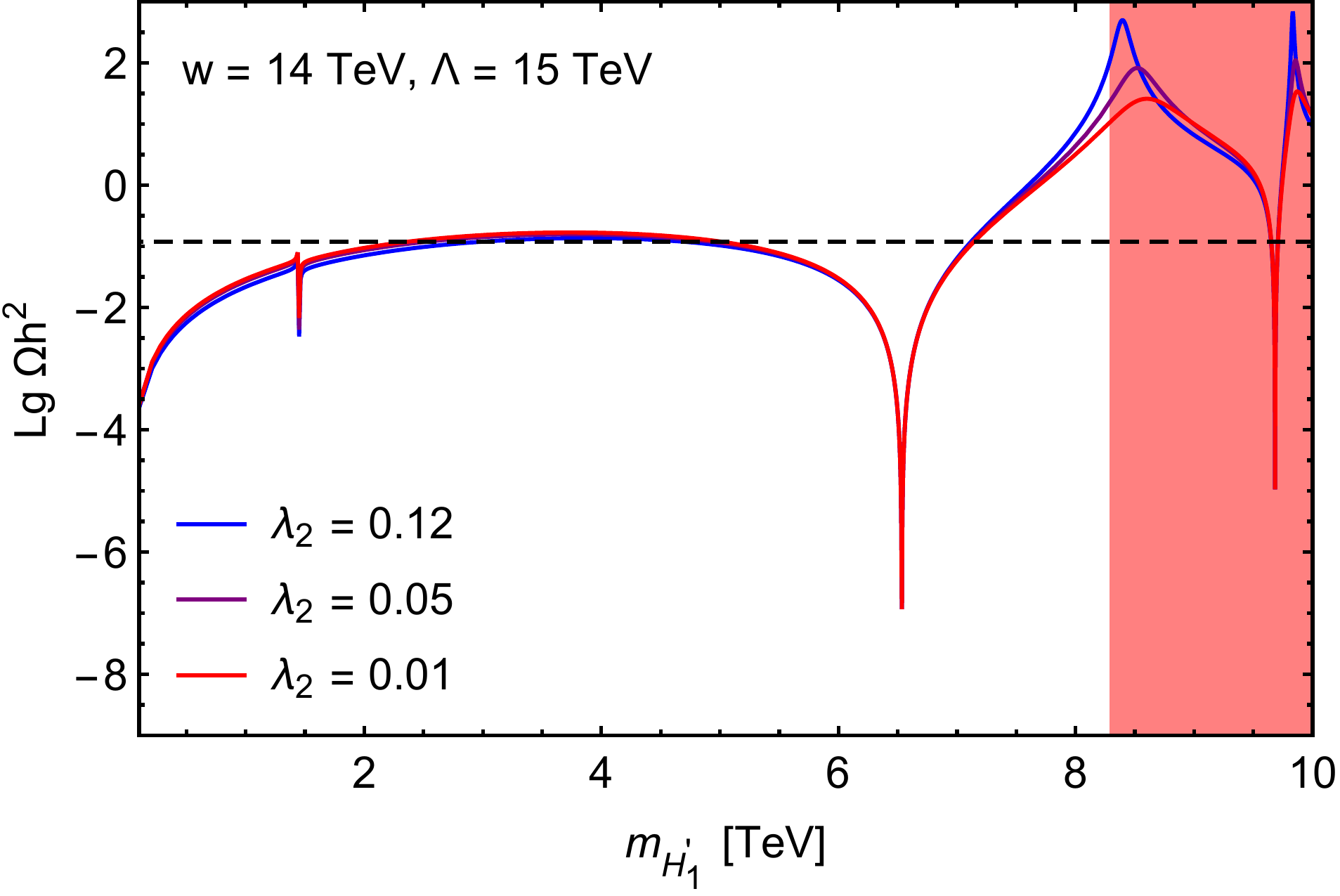}
	\caption{\label{rdesca1} The relic density of the scalar candidate as a function of the DM mass according to the several choices of $w,\Lambda$, where the DM unstable region (light red) is the band on the right side of each panel.}	
\end{center}
\end{figure}

In Figure \ref{rdesca1}, we depict the relic density as a function of DM mass which all vary upon the variation of $f_2$ and simultaneously show the DM unstable regime (light red, when $H'_1$ is not the lightest $W_P$-odd field) according to the several selections of the new physics scales $w$ and $\Lambda$. These selections are consistent with the constraints from the $\rho$-parameter, the LHC dilepton, as well as the charged lepton flavor violating processes and nonstandard neutrino interactions, as studied in \cite{Huong:2019vej}. Additionally, throughout this paper the choice $\Lambda\sim w$ would characterize the strength of the $SU(3)_L$ breaking down to $SU(2)_L$, as assumed from the model setup. The viable DM mass regime is given below the correct abundance $\Om h^2<0.12$ and before the DM unstable regime. We see that each density curve contains three resonances, corresponding to $m_{H'_1} = m_{H_2}/2$, $m_{H'_1} = m_{H_3}/2$ and $m_{H'_1} = m_{H_4}/2$, where the relic density is largely decreased. The variation of $\la_2$ slightly separates the relic density and this is also valid for the whole viable range of $\la_2\sim 0$--0.7, since $\la_1>0$. When $w,\La$ are large as in the right panel, the DM mass upper bound is increased but appearing an excluded intermediate region according to $\Om h^2>0.12$. 

\begin{figure}[!h]
\begin{center}
	\includegraphics[scale=0.4]{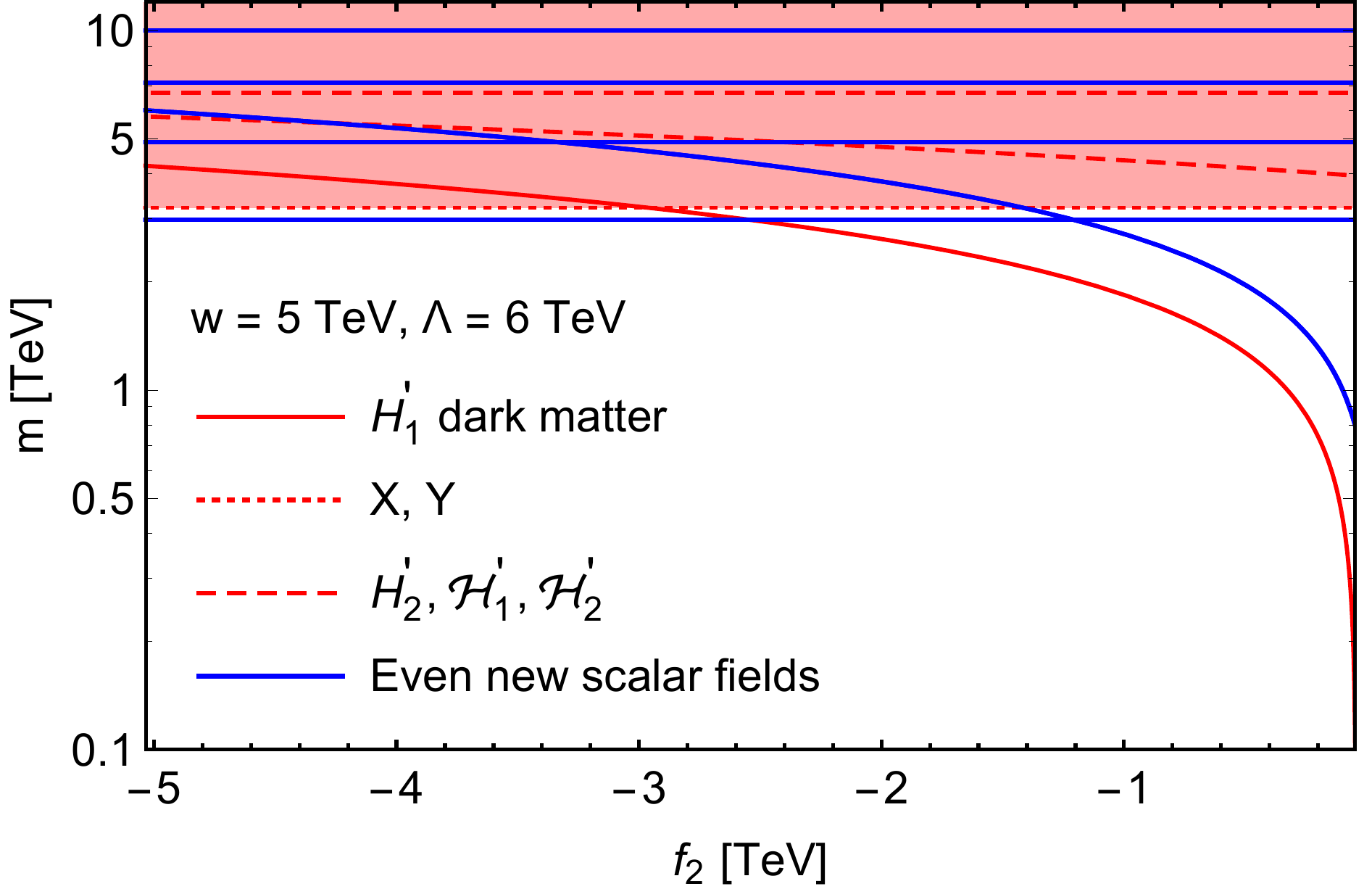}
	\includegraphics[scale=0.4]{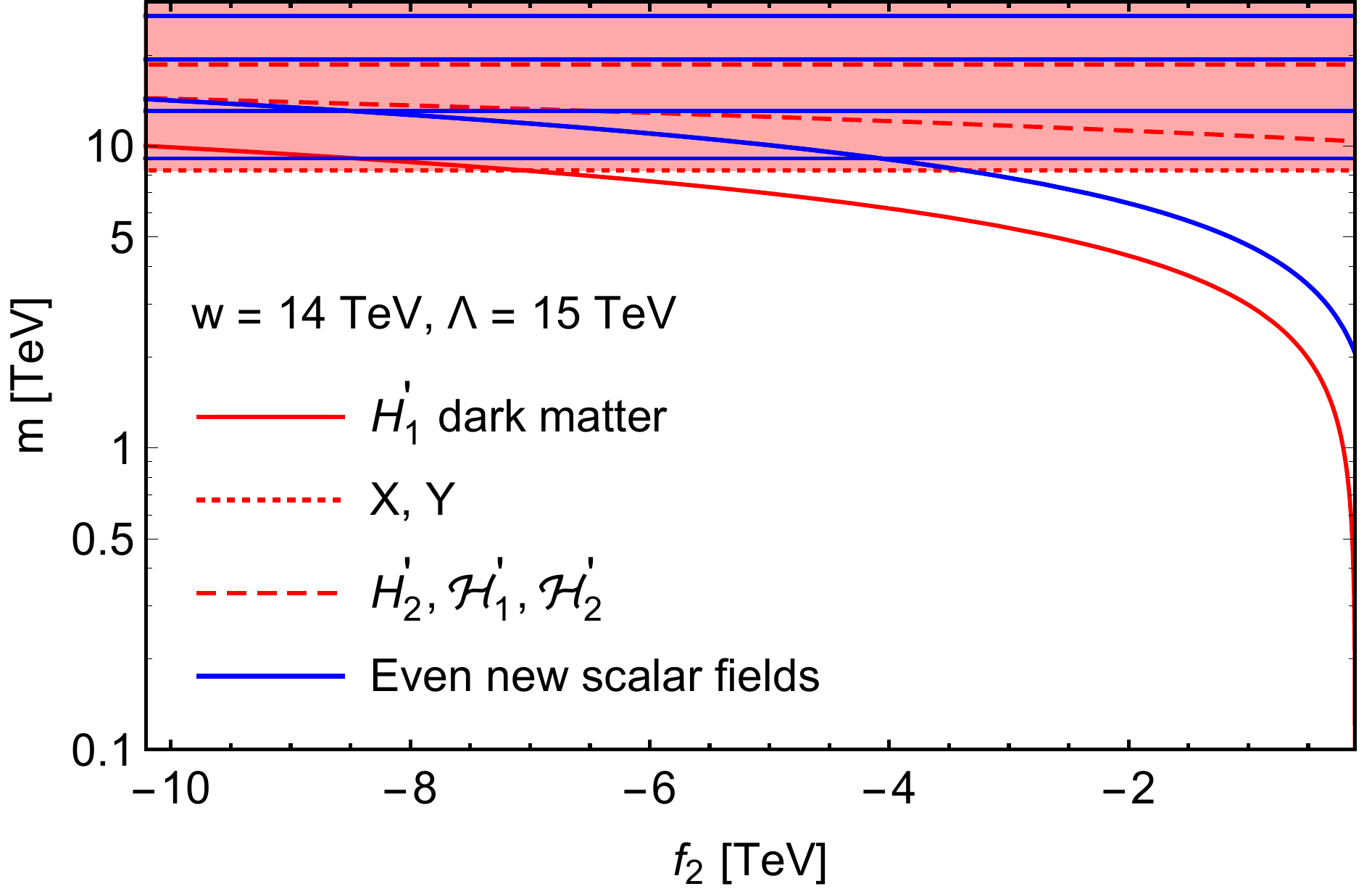}
	\caption{\label{rdesca2} The masses of the new scalar fields and the odd gauge bosons as functions of the parameter $f_2$ according to the several choices of $w,\Lambda$, where the dark field lines are put as red and the DM stable regime (i.e., below the light red background) is bounded by the $X,Y$ line.}	
\end{center}
\end{figure}
With the benchmark choices in Eq. (\ref{pafosca}), the DM mass variation in Fig. \ref{rdesca1} is exclusively obtained by varying $f_2$, as mentioned. But, such variation of $f_2$ also leads to the changes of the masses of other fields, which are described in Fig. \ref{rdesca2}. Here we do not explicitly label the lines of the same field type for simplicity. Note also that the $X,Y$ gauge bosons are almost degenerated in mass. Obviously, all the new scalar fields in the model except the dark matter obtain mass values safely above $\mathcal{O}(1)$ TeV. There is a range of $f_2$ where $H'_1$ is the lightest $W_P$-odd field, set by the $X,Y$ mass, as imposed to Fig. \ref{rdesca1}. Note that in the unstable region of Fig. \ref{rdesca1}, the DM scalar proceeds to decaying into $X,Y$ bosons through channels, $H'_1\rightarrow YH_1, YZ, XW$, derived by the corresponding vertices in Appendix \ref{coupscalars}, which would be kinematically suppressed. Last, the dark fermions $\xi,E,U$ have masses proportional to the relevant Yukawa couplings in Eq. (\ref{addttba}) which can easily be chosen, such that these fermions are heavier than the scalar DM. So we need not refer to this case here.

\begin{figure}[!h]
\begin{center}
	\includegraphics[scale=0.4]{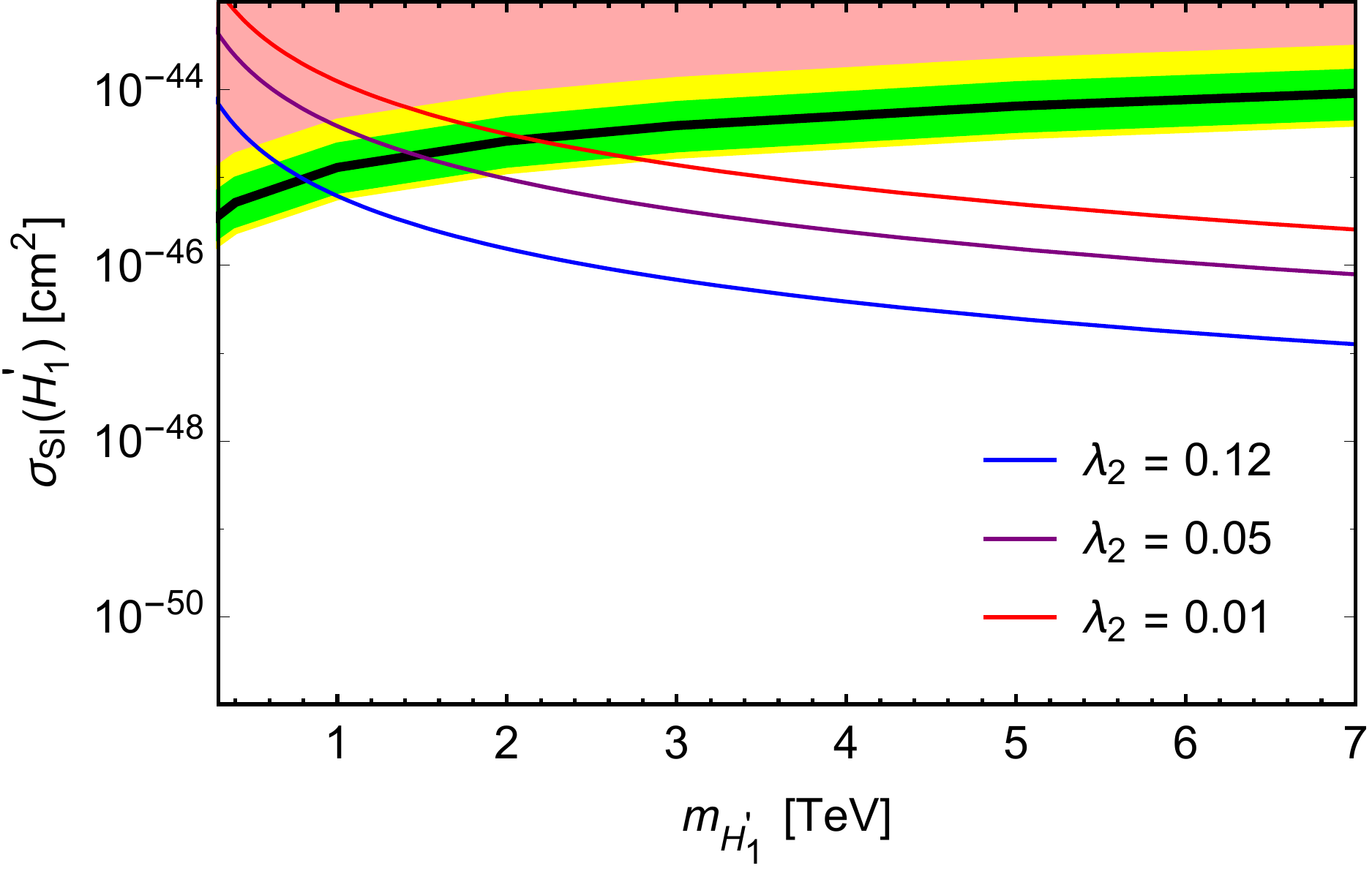}
	\caption{\label{ddesca} The SI DM-nucleon scattering cross-section limit as a function of DM mass according to the several choices of $\lambda_2$, where the excluded region (light red) lies above the experimental (green and yellow) bands.}	
\end{center}
\end{figure}

The SI DM-nucleon cross-section (\ref{t1ln321}) depends only on the DM mass $m_{H'_1}$ and the scalar coupling $\la_2$, where the remaining parameters $u,v,\la_6$ were fixed as mentioned. The choice, change, or variation of all the relevant parameters in $m_{H'_1}$, such as $w,\La,f_2$ and so on, effectively modifies only the DM mass. Hence, although we have several cases for parameter values, it all leads to the cross-section dependence only in terms of $m_{H'_1},\la_2$, yielding a common result. In Figure \ref{ddesca}, we depict the SI scattering cross-section of $H'_1$ on the nucleon according to the above choices of $\lambda_2$ as well as show the experimental bounds from XENON1T \cite{Aprile:2018dbl}, where the black line is the SI WIMP-nucleon cross-section limit at 90\% confidence level, while the green and yellow bands are the $1\sigma$ and $2\sigma$ sensitivities, respectively. We see that the scalar DM mass below around 500 GeV is excluded by the direct detection experiment for $\la_2=0.12$. Hence, in this case the viable scalar DM mass regime is around 500 GeV to a few TeV (cf. Figure \ref{rdesca1}). However, the SI DM cross-section is quite separated for the variation of $\la_2$. In the latter cases for $\la_2=0.05$ and 0.01, the respective lower bounds of $H'_1$ mass are above 1 TeV. If $\la_2$ is bigger, i.e. $\la_2>0.12$, the DM mass is close to the weak scale, and in this case the DM would be subject to the electroweak precision test and collider bounds, which is not studied in this work. Additionally, when $\la_2$ is smaller than 0.01, the viable DM mass region is narrow as constrained by the relic density and the unstable regime.

\subsubsection{Varying $w,\La$ while fixing relevant couplings}

\begin{figure}[!h]
\begin{center}
	\includegraphics[scale=0.4]{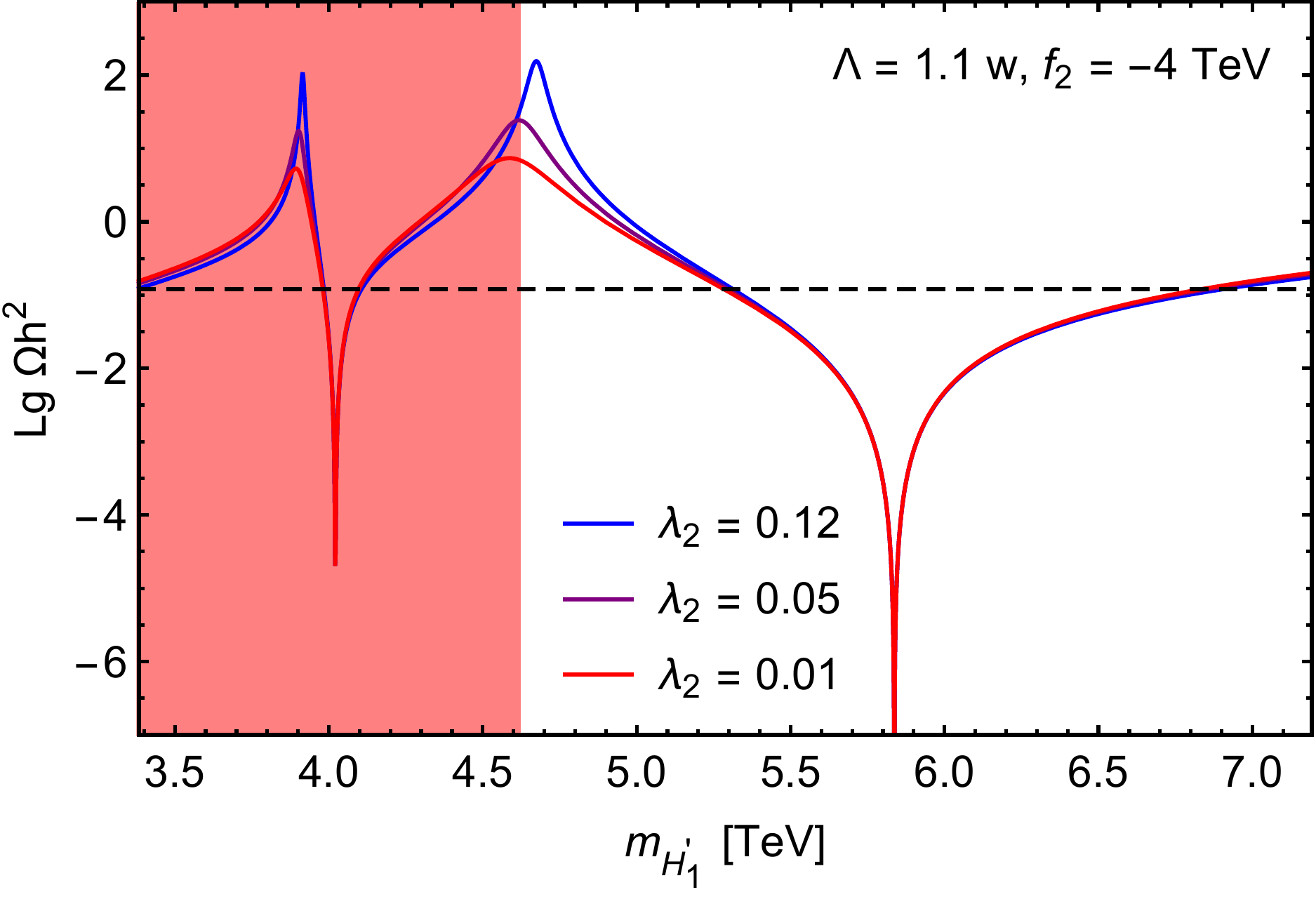}
	\includegraphics[scale=0.4]{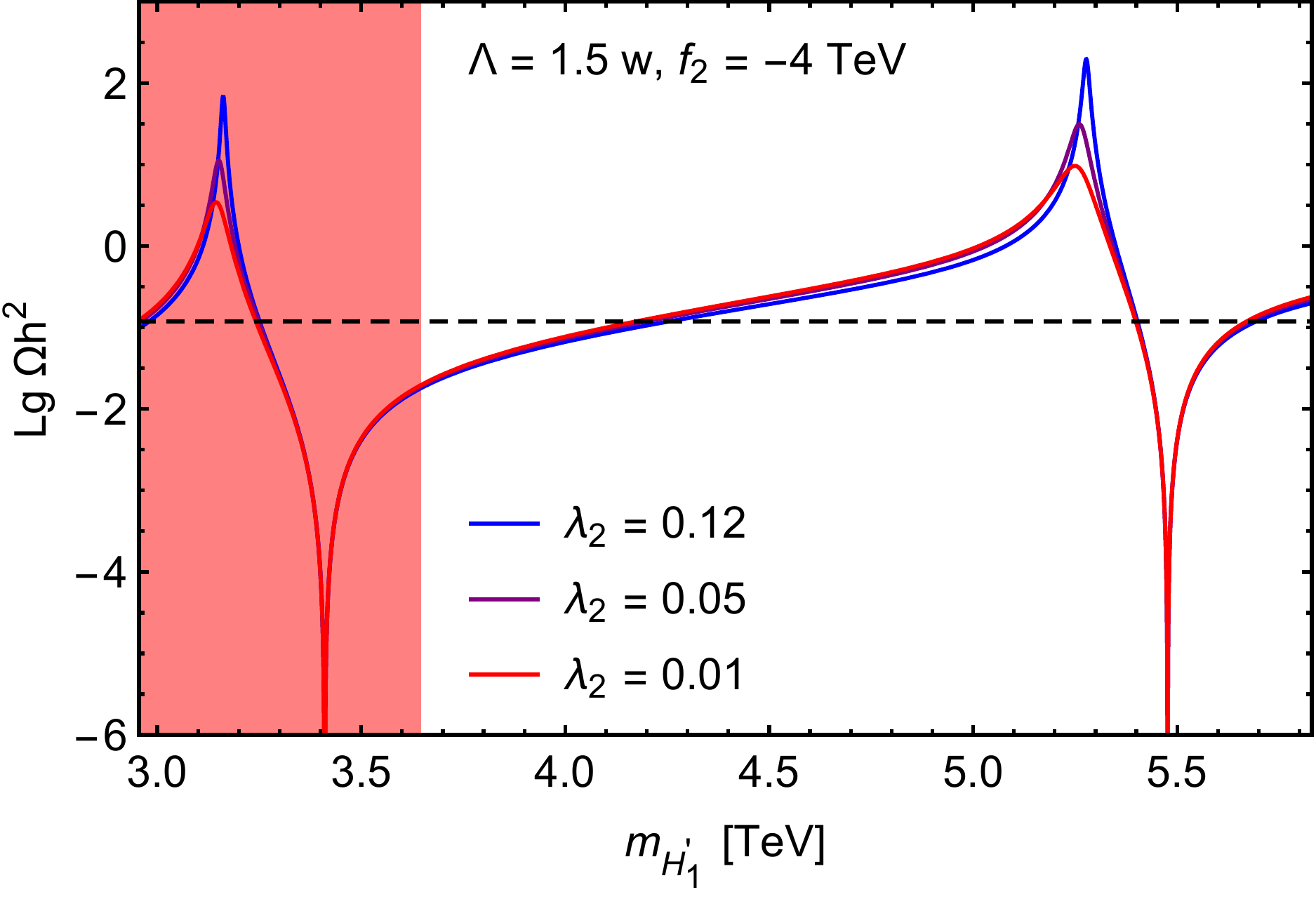}
	\caption{\label{rdesca3} The relic density of the scalar candidate as a function of the DM mass, which all vary according to the variation of $w$ and $\Lambda$, but with several relations as fixed, where the DM unstable region (light red) is the band on the left side of each panel.}	
\end{center}
\end{figure}

\begin{figure}[!h]
\begin{center}
	\includegraphics[scale=0.4]{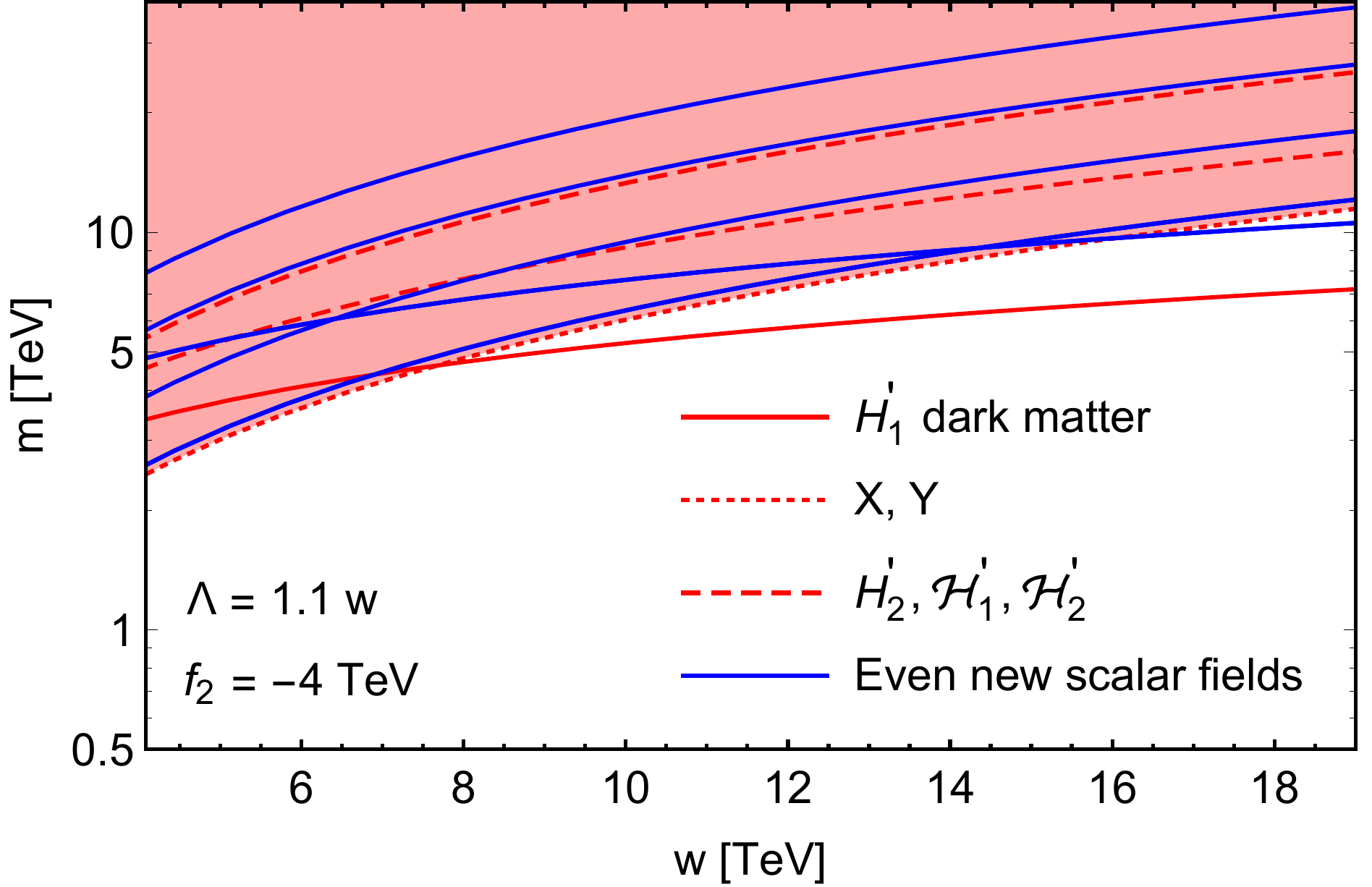}
	\includegraphics[scale=0.4]{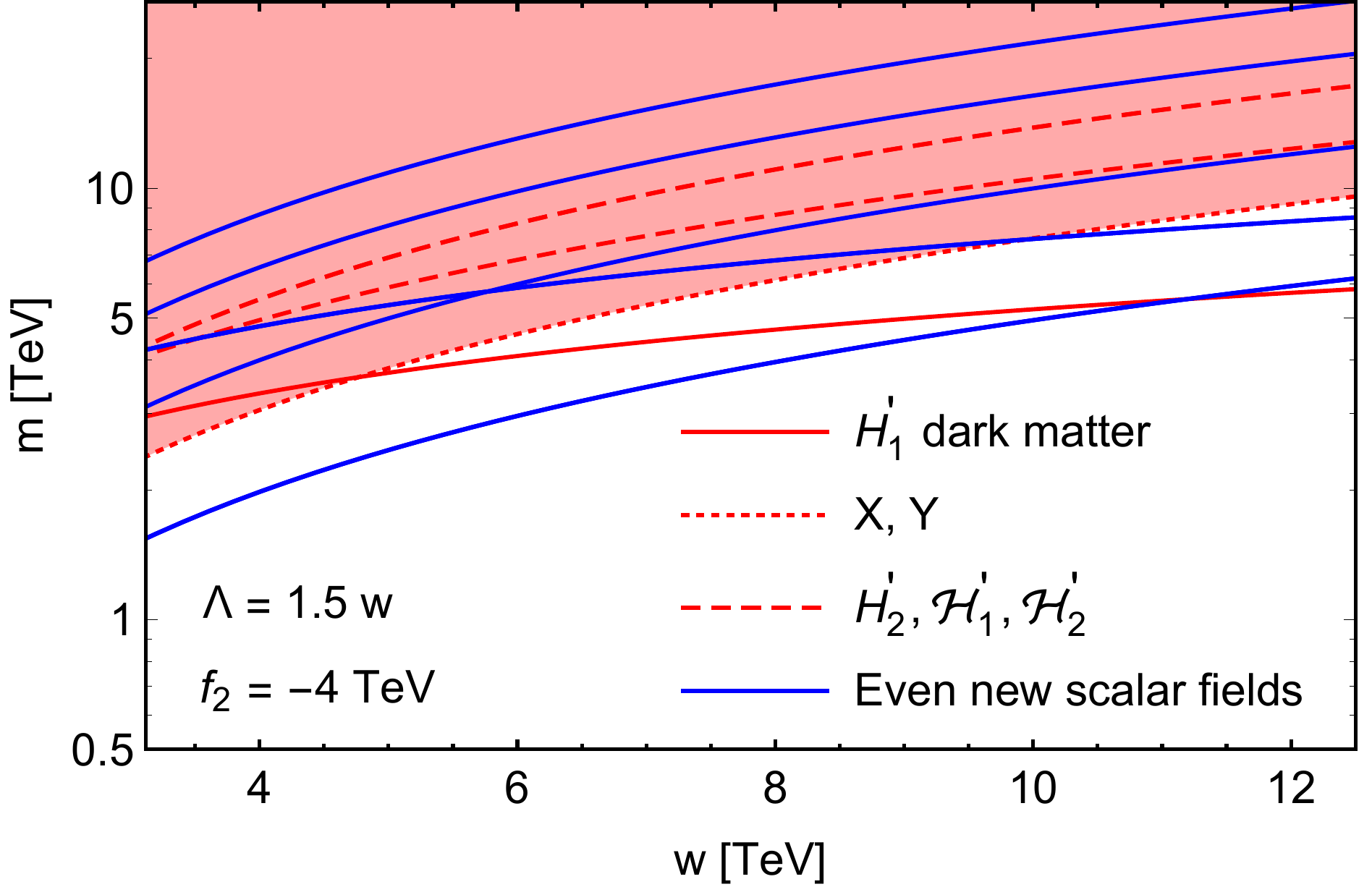}
	\caption{\label{rdesca4} The masses of the new scalar fields and the odd gauge bosons as functions of $w$ according to the several relations between $w$ and $\Lambda$, where the dark field lines are put as red and the DM unstable regime is input as light red background, which is bounded by $X,Y$ line.}	
\end{center}
\end{figure}

Since the new physics scales $w$ and $\Lambda$ govern all the DM  annihilation channels, it is interesting to recast the above investigation in which $w$ and $\Lambda$ are varied instead of $f_2$. For this aim, we choose $f_2=-4$ TeV, without loss of generality. As mentioned, $w,\La$ commonly set the strength of the $SU(3)_L$ breaking, it is suitably to impose $w\sim \La$, while both these parameters are simultaneously run from a lower bound as supplied in Eq. (\ref{lddabc}). We also vary $\la_2$ while fixing the values of the other parameters as in the previous case. The results of the relic density and the mass variations are shown in Fig. \ref{rdesca3} and Fig. \ref{rdesca4}, respectively. 

As we see from Fig. \ref{rdesca4}, all the new particle masses significantly change upon the new physics scales, apposite to the previous case as $f_2$ varies. Additionally, the DM scalar is stabilized if $w>7.75$ TeV and 4.75 TeV corresponding to the cases in the left and right panels, respectively. This translates to the unstable regions as lower bounds on the DM mass according to the left and right panels of Fig. \ref{rdesca3}, unlike the previous case. With the choice of the parameters, there are only two resonances in the relic density as seen in Fig. \ref{rdesca3} set by the $H_{3,4}$ masses, i.e. $m_{H'_1}=\fr 1 2 m_{H_{3}}$ and $m_{H'_1}=\fr 1 2 m_{H_{4}}$, respectively. This can also be realized from Fig. \ref{rdesca4} by comparing relevant masses. The splittings in the relic density and the SI DM cross-section according to $\la_2$ are similar to the previous case. Here note that the SI DM cross-section remains the same previous case, since it depends only on the DM mass and $\la_2$.       

That said, the viable scalar DM mass regime is now either from 5.25 TeV to 6.85 TeV according to $\Lambda=1.1 w$ or from 3.65 TeV to 4.2 TeV and from 5.4 TeV to 5.67 TeV according to $\Lambda=1.5 w$, as bounded by the unstable region and the relic density $\Om h^2<0.12$. Such ranges are appropriate to the SI DM cross-section bounds.

\subsection{DM as a triplet fermion}

The $\xi$ triplet components have a degenerate mass, \be m_\xi=-\sqrt{2}h^\xi \La,\ee at the tree level, as induced by the VEV of the scalar sextet. But, the loop effects of gauge bosons can make $\xi^\pm$ mass larger than $\xi^0$ mass by an amount, \be m_{\xi^\pm}-m_{\xi^0}=166\ \mathrm{MeV},\ee as shown in \cite{Cirelli:2005uq}. Hence, $\xi^0$ is first regarded as the lightest of the triplet components. We further assume that $\xi^0$ is the lightest particle among all of the other $W_P$-odd particles and thus $\xi^0$ is responsible for the DM candidate. This scenario was briefly discussed in \cite{Huong:2019vej}, in which the field $\xi^0$ yielded the correct abundance and satisfied the direct detection bounds, provided that it had a mass $m_{\xi^0}\simeq 2.86$ TeV. Here in the present work, we will explore the full viable mass region of $\xi^0$ and investigate physical density resonances which are crucial for the experimental detections.

\begin{figure}[h]
	\centering
	\includegraphics[scale=0.8]{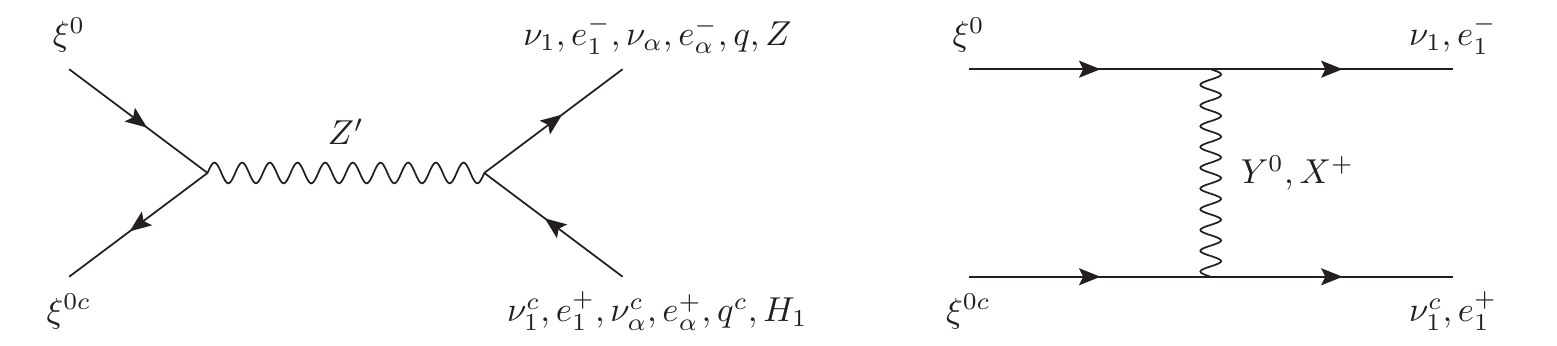}
	\caption{Dominant contributions to the fermion DM pair annihilation into the SM particles.}	\label{annixi0}
\end{figure}

It is obvious that the DM candidates mainly annihilate to the SM particles. The dominant channels for the fermion DM pair annihilation $\xi^0$ to the SM particles are presented in Figure \ref{annixi0}. 

The thermal average annihilation cross-section times the relative velocity is approximated as follows 
\bea
\langle\sigma v\rangle_{\xi^0\xi^{0c}\to\mathrm{SM\ SM}} &=& \frac{g^4m^2_{\xi^0}}{\pi c^4_W}\left\{\left[g^{Z'}_V(\xi^0)\right]^2\left[ \sum_{f} N_C(f)\frac{[g^{Z'}_V(f)]^2+[g^{Z'}_A(f)]^2}{16(4m^2_{\xi^0}-m^2_{Z'})^2} \right.\right.\crn
&&+\left. \frac{g^2_{Z'ZH_1}c^2_W}{32g^2m^2_Z (4m^2_{\xi^0}-m^2_{Z'})^2} \right] + \left[\frac{c^4_W}{(m^2_{\xi^0}+m^2_Y)^2}\right.\crn
&&-\left.\left.\frac{g^{Z'}_V(\xi^0) g^{Z'}_V(\nu_1)c^2_W}{2(m^2_{\xi^0}+m^2_Y)(4m^2_{\xi^0}-m^2_{Z'})}+(\nu_1\leftrightarrow e_1, m_Y\leftrightarrow m_X)\right] \right\},
\eea
where $f$ denotes the SM fermions and the masses of the new gauge bosons are assumed to be much larger than the masses of the SM fermions. Note that $\xi^0$ can also annihilate to the SM Higgs boson via a $s$-channel diagram, exchanged by the new Higgs portal, $H_{3,4}$, because $H_{3,4}$ couples to $\xi^0$ via the coupling $h^\xi\sim m_\xi/\La$ and to $H_1$ via a coupling in the form of $\bar{\la}\times (w,\La)$, where $\bar{\la}$ is a combination of the known scalar self-couplings, $\la$'s. This contribution may become evident with the interesting resonances at $m_{\xi^0}=m_{H_{3,4}}/2$ in the relic density, similar to the scalar DM case, as studied above. In the present case, we are most favored in investigating the gauge portals, which are not significant in the scalar DM case. For this aim, the $H_{3,4}$ contributions to the total annihilation cross-section can directly be suppressed, as omitted, by assuming that $\bar{\la}$ is radically smaller than the gauge coupling constant, $g$.   

Note that $\xi^0$ does not interact with $Z$ at the effective limit $u,v\ll w,\La$. To investigate the direct detection of the field $\xi^0$ via the SI scattering on nuclei, we write the effective Lagrangian that describes the interactions of $\xi^0$ with fundamental level quarks, induced through the $t$-channel exchange of the field $Z'$, such as
\be \mathcal{L}^{\mathrm{eff}}_{\xi^0-\mathrm{quark}} = \frac{g^2}{4c^2_W m^2_{Z'}}\bar{\xi}^0\gamma^\mu g^{Z'}_V(\xi^0)(1-\gamma_5)\xi^0\bar{q}\gamma_\mu [g^{Z'}_V (q)-g^{Z'}_A (q)\gamma_5]q.\ee
Thus, we obtain the SI cross-section for the scattering of the $\xi^0$ on a target nucleus \cite{Belanger:2008sj},
\be \sigma^{\mathrm{SI}}_{\xi^0 N} = \frac{g^4 m^2_{\xi^0 N}}{16\pi c^4_W m^4_{Z'}} [g^{Z'}_V(\xi^0)]^2 \left|g^{Z'}_V (u) (Z+A)+g^{Z'}_V (d) (2A-Z)\right|^2,\ee
where $m_{\xi^0 N}\simeq m_N$. 

\begin{figure}[!h]
\begin{center}
	\includegraphics[scale=0.4]{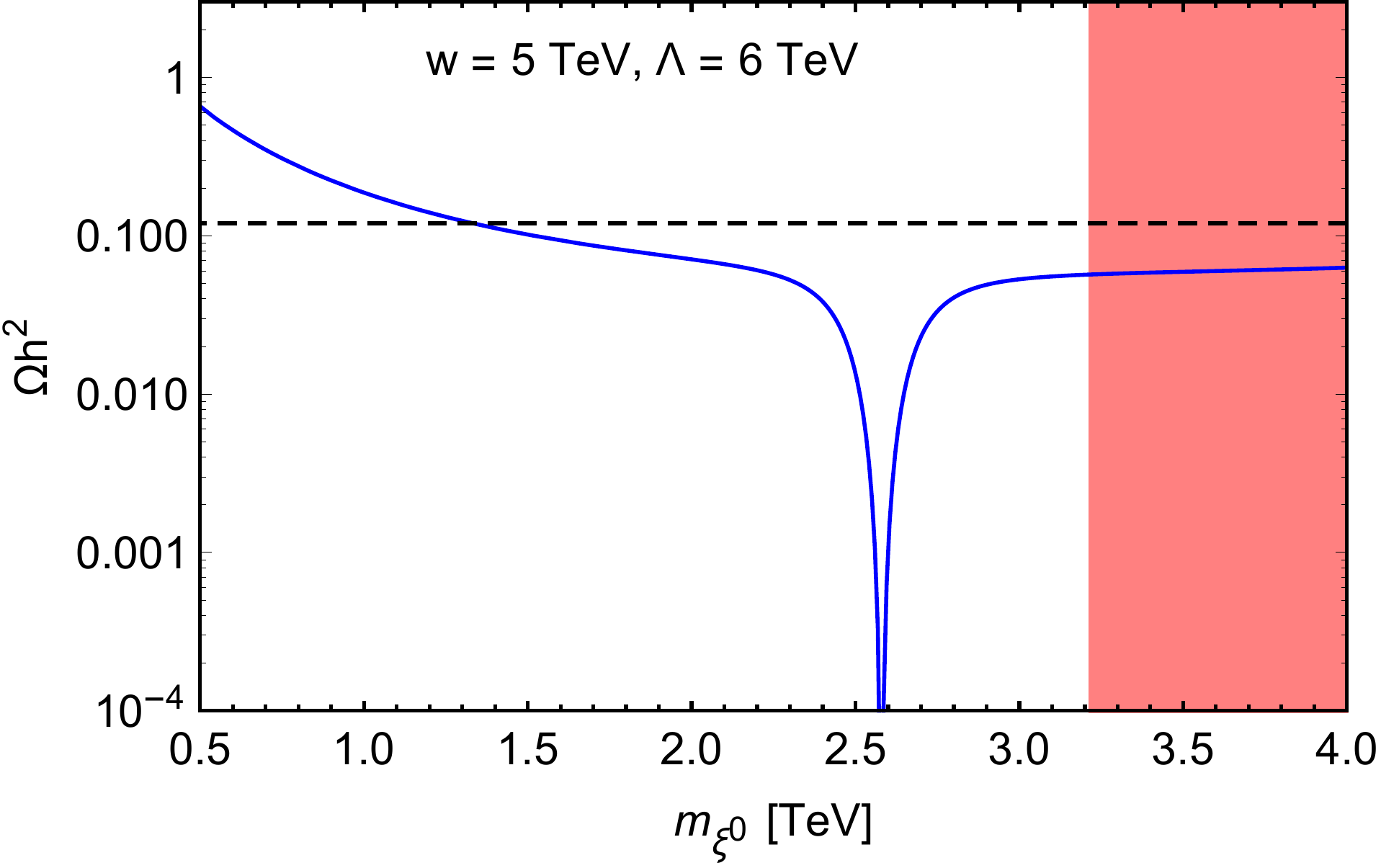}
	\includegraphics[scale=0.4]{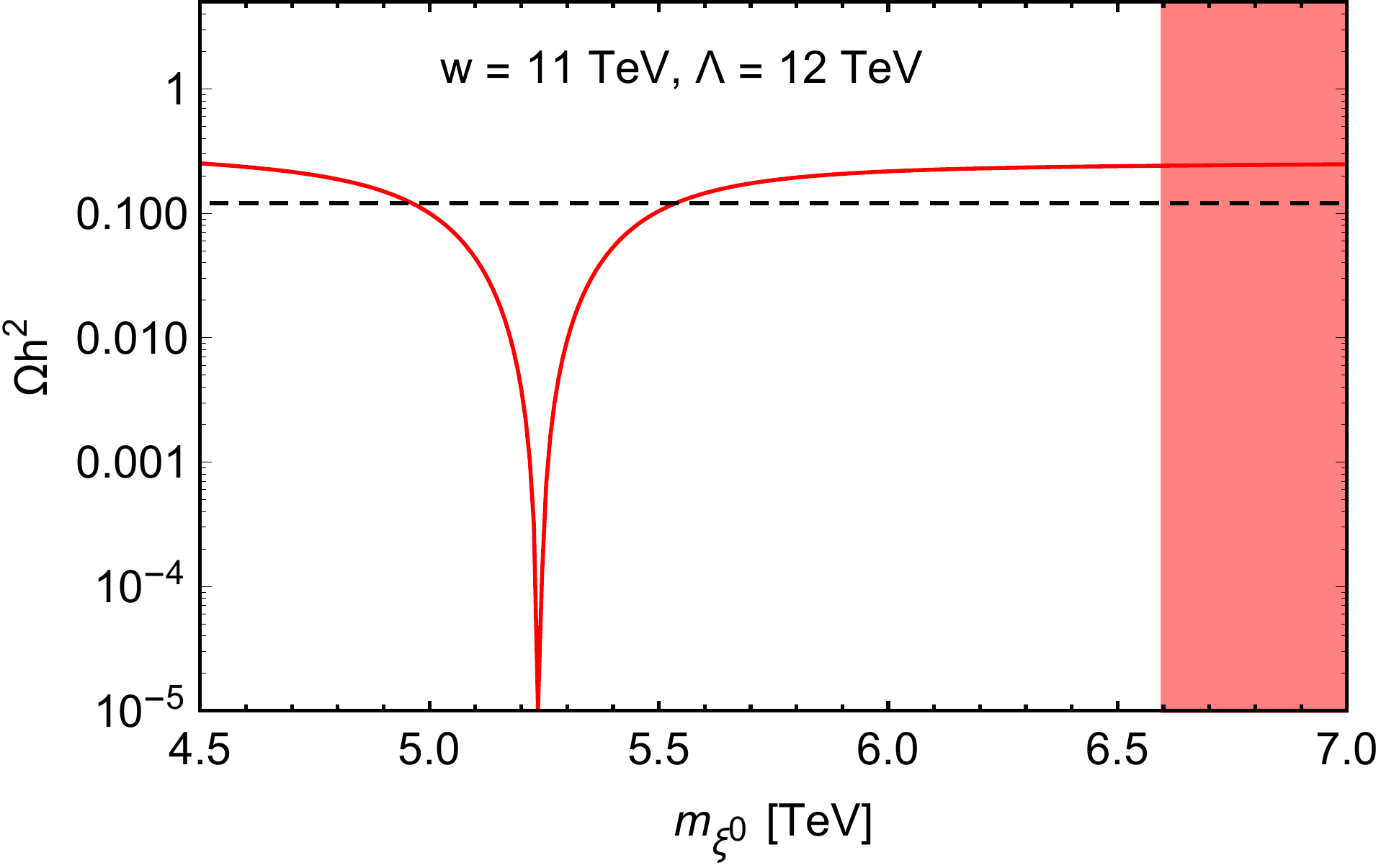}
	\caption{\label{trdfer} The relic density of the fermion candidate as a function of its mass according to the several choices of $w,\Lambda$, where the DM unstable region is input as light red.}	
\end{center}
\end{figure}

\begin{figure}[!h]
\begin{center}
	\includegraphics[scale=0.4]{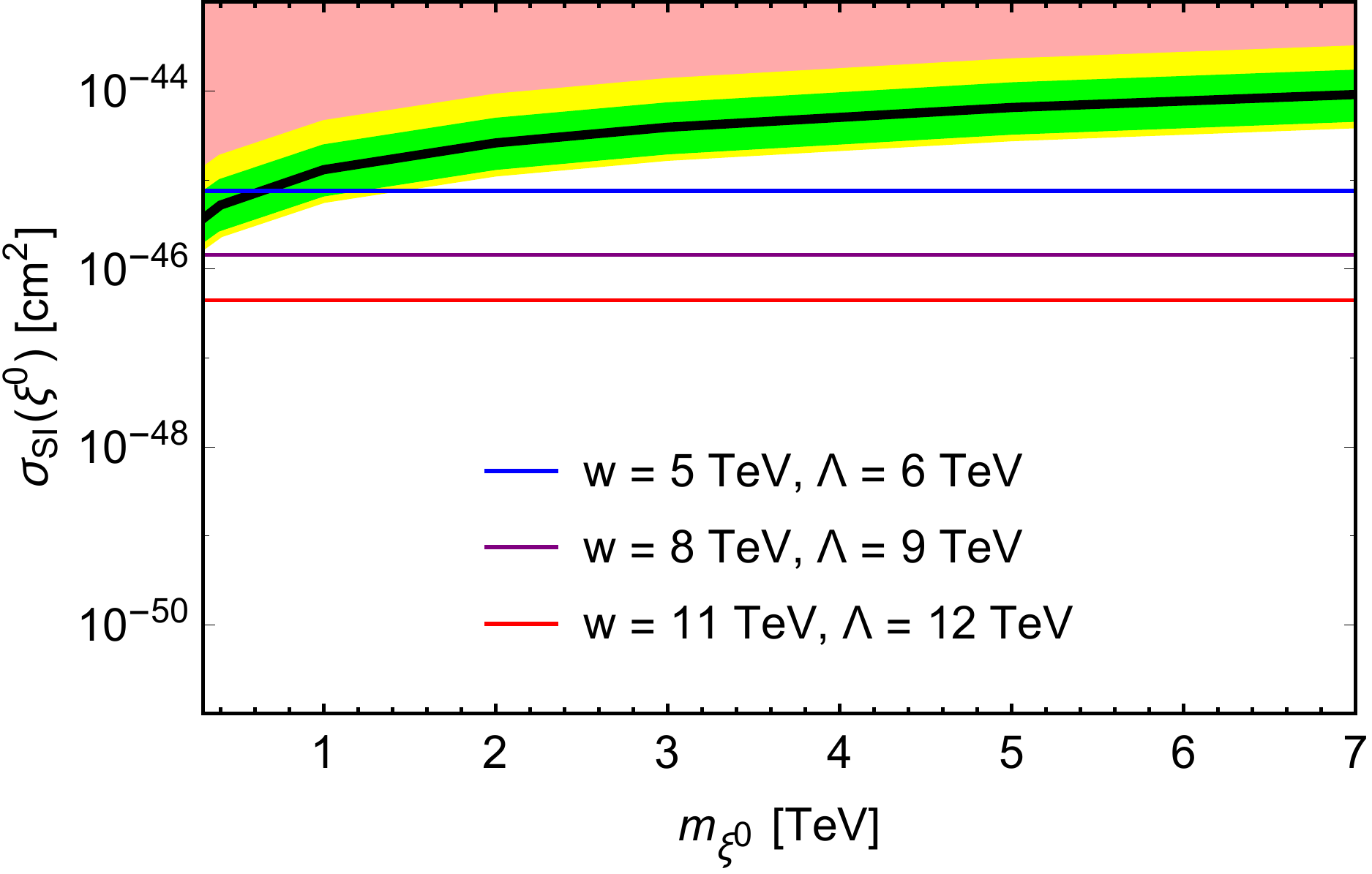}
	\caption{\label{ddefer} The SI DM-nucleon scattering cross-section limit as a function of DM mass according to the several choices of $w, \Lambda$, where the excluded region is input as light red.}	
\end{center}
\end{figure}

For numerical computation in this subsection, we take $m_Z\simeq 91.187$ GeV and relevant parameters given in (\ref{pafoscaddttt}). In Figure \ref{trdfer}, we plot the relic density of the DM as a function of its mass according to the several choices of $w$ and $\Lambda$. We see that each density curve always contain a quite narrow resonance at $m_{\xi^0}=m_{Z'}/2$, at which the relic density is substantially reduced, tending to zero. Whereas, outside the resonance region, the relic density slowly increases. Since the gauge couplings are fixed and that the $Z$-$Z'$ mixing angle is small, the relic density only depends on $m_{\xi^0}$ and $w,\La$ through $m_{X,Y,Z'}$. It is proportional to $m^4_{X,Y,Z'}/m^2_{\xi^0}$ for $m_{\xi^0}$ close to the weak scale, while it is proportional to $m^2_{\xi^0}$ for $m_{\xi^0}$ much beyond $m_{X,Y,Z'}$. Because the range of the DM mass considered in Fig. \ref{trdfer} is narrow, the corresponding relic density outside the resonance region is weakly changed as scaled by the resonance mass $m_{Z'}$. To see a significant change, we make an estimation, $\Om h^2=2.8$, 0.7, and 3.75 for $m_{\xi^0}=1$, 20, and 50 TeV, respectively, according to the right panel. Whereas, $\Om h^2=0.6$ and 3.6 correspond to $m_{\xi^0}=20$ and 50 TeV, respectively, according to the left panel. In Figure \ref{ddefer}, we plot the SI scattering cross-section limit as a function of the DM mass according to the above choices. The limits are in good agreement the constraint from XENON1T \cite{Aprile:2018dbl}. Combining the results in Fig. \ref{trdfer} and Fig. \ref{ddefer}, the viable DM mass regime is as follows: $m_{\xi^0}=1.3$--3.2 TeV for $w=5$ TeV, $\Lambda=6$ TeV; $m_{\xi^0}=3.5$--4.4 TeV for $w=8$ TeV, $\Lambda=9$ TeV; and $m_{\xi^0}=4.9$--5.5 TeV for $w=11$ TeV, $\Lambda=12$ TeV. Here the middle one is set by the relic density with a plot similar to the right panel in Fig. \ref{trdfer}, which was not displayed.

Last, but not least, the scalar sextet ($S$) and triplet ($\chi$) are decomposed as \be S\sim 6=3\oplus 2\oplus 1,\hs \chi\sim 3=2\oplus 1,\ee under the SM $SU(2)_L$ symmetry at the effective limit. Here $S_{33}$ and $\chi_3$ transform as the $SU(2)_L$ singlets respectively, hence they can develop the VEVs, $\langle S_{33}\rangle =\La/\sqrt{2}$ and $\langle \chi_{3}\rangle = w/\sqrt{2}$, for breaking $SU(3)_L$ down to $SU(2)_L$ as conserved. At this stage, $w,\La$ contribute to all the new particle masses, as seen in the new gauge and Higgs bosons above. Additionally, $w$ gives exotic quark mass, while $\La$ provides $\xi$ mass, and both $w,\La$ supply $E$ mass. All that implies a similar role between $w,\La$ as mutually contributing to the new physics and interacting of their fields with the SM. Although we have chosen $\La>w$ in interpreting the results, we reexamined that an opposite choice, $w>\La$, or including both cases, but always ensuring $w\sim \La$, lead to the same physics.

\section{\label{conl}Conclusion}

In this work, the fully flipped 3-3-1-1 model has been interpreted. The scalar sector has explicitly been diagonalized, yielding the appropriate particle spectrum. We have shown that the model naturally provides the two DM candidates, a singlet scalar and a triplet fermion, with the masses in TeV regime. We have determined the physical resonances in the dark matter relic density, which are set by the new neutral gauge boson and the new neutral Higgs boson according to the fermion and scalar densities, respectively. A further experimental search for these resonances is crucial to probe the existence of dark matter.    

\section*{Acknowledgments}

This research is funded by Vietnam National Foundation for Science and Technology Development (NAFOSTED) under grant number 103.01-2019.353.

\appendix
\section{\label{coupscalars} Couplings of fermions and scalars with gauge bosons}
This appendix is devoted to determine all the couplings of fermions and scalars with gauge bosons, given throughout Tables \ref{Zgiffermion}, \ref{1NG2STab1}, \ref{1CG2STab1}, \ref{1S2GTab1}, \ref{1NG1CG2STab1}, \ref{1NG1CG2STab2}, \ref{1NG1CG2STab3}, \ref{2CG2STab1}, \ref{2CG2STab2}, \ref{2NG2STab1}, and \ref{2NG2STab2}, according to the various types of interactions. Here, we have frequently utilized the notations, $A \overlr{\partial} B \equiv  A(\partial B)-(\partial A) B$ and $\Upsilon \equiv \sqrt{(4\La^2+w^2)/(u^2+v^2)}$. While the form of the fermion and gauge boson interactions is explicitly displayed in the body text, let us remind the reader that each scalar and gauge boson interaction is directly determined as its vertex times coupling; neither extra imaginary unit nor derivatives with respect to the vertex fields as set for the Feynman rules are supplied. 

\begin{table}[t]
\bc
\begin{tabular}{lcc}
\hline\hline
$f$ & $g^{Z}_V(f)$ & $g^{Z}_A(f)$ \\\hline 
$\nu_1$ & $\frac{s_\varphi c_{2W}+c_\varphi\sqrt{1+2c_{2W}}}{2\sqrt{1+2c_{2W}}}$ & $\frac{s_\varphi c_{2W}+c_\varphi\sqrt{1+2c_{2W}}}{2\sqrt{1+2c_{2W}}}$ \\
$\nu_\alpha$ & $-\frac{s_\varphi-c_\varphi\sqrt{1+2c_{2W}}}{2\sqrt{1+2c_{2W}}}$ & $-\frac{s_\varphi-c_\varphi\sqrt{1+2c_{2W}}}{2\sqrt{1+2c_{2W}}}$ \\
$e_1$ & $\frac{(s_\varphi-c_\varphi\sqrt{1+2c_{2W}})(2c_{2W}-1)}{2\sqrt{1+2c_{2W}}}$ & $\frac{s_\varphi-c_\varphi\sqrt{1+2c_{2W}}}{2\sqrt{1+2c_{2W}}}$ \\ 
$e_\al$ & $\frac{s_\varphi(c_{2W}-2)-c_\varphi\sqrt{1+2c_{2W}}(2c_{2W}-1)}{2\sqrt{1+2c_{2W}}}$ & $-\frac{s_\varphi c_{2W}+c_\varphi\sqrt{1+2c_{2W}}}{2\sqrt{1+2c_{2W}}}$ \\ 
$u_a$ & $\frac{s_\varphi(4-c_{2W})+c_\varphi\sqrt{1+2c_{2W}}(4c_{2W}-1)}{6\sqrt{1+2c_{2W}}}$ & $\frac{s_\varphi c_{2W}+c_\varphi\sqrt{1+2c_{2W}}}{2\sqrt{1+2c_{2W}}}$ \\
$d_a$ &$\frac{(s_\varphi-c_\varphi \sqrt{1+2c_{2W}}) \sqrt{1+2c_{2W}}}{6}$ &  $\frac{s_\varphi-c_\varphi\sqrt{1+2c_{2W}}}{2\sqrt{1+2c_{2W}}}$\\
$\xi^0$ & $-\frac{s_\varphi c_{W}^2}{\sqrt{1+2c_{2W}}}$ &  $-\frac{s_\varphi c_{W}^2}{\sqrt{1+2c_{2W}}}$ \\
$\xi$ & $2c_\varphi c^2_W$ &  $-\frac{2s_\varphi c_{W}^2}{\sqrt{1+2c_{2W}}}$ \\
$E_1$ & $\frac{2s_\varphi c_{2W}+c_\varphi\sqrt{1+2c_{2W}}(1-c_{2W})}{\sqrt{1+2c_{2W}}}$ & $\frac{2s_\varphi c_{W}^2}{\sqrt{1+2c_{2W}}}$\\
$E_\al$ & $\frac{s_\varphi(3c_{2W}-1)-2c_\varphi\sqrt{1+2c_{2W}}(c_{2W}-1)}{2\sqrt{1+2c_{2W}}}$ & $\frac{s_\varphi c_{W}^2}{\sqrt{1+2c_{2W}}}$\\
$U_a$ & $\frac{s_\varphi(1-7c_{2W})+4c_\varphi\sqrt{1+2c_{2W}}(c_{2W}-1)}{6\sqrt{1+2c_{2W}}}$ & $-\frac{s_\varphi c^2_W}{\sqrt{1+2c_{2W}}}$ \\
\hline\hline
\end{tabular}
\caption{\label{Zgiffermion}The couplings of $Z$ with fermions ($f\ne\nu_R$), whereas those for $Z'$ are obtained by replacement: $c_\varphi\to s_\varphi$ and $s_\varphi\to -c_\varphi$.}
\ec
\end{table}

\begin{table}[!h]
\begin{ruledtabular}
\begin{tabular}{lccc}
Vertex & Coupling & Vertex & Coupling\\ \hline
$A_\mu H_{11}^{++} \overlr{\partial}^\mu H_{11}^{--}$ & $2ig s_W$ & $A_\mu H_{12}^{+} \overlr{\partial}^\mu H_{12}^{-}$ & $ig s_W$ \\
$A_\mu H_{1}^{+} \overlr{\partial}^\mu H_{1}^{-}$ & $ig s_W$ & $A_\mu \mathcal{H}'^{+}_{1} \overlr{\partial}^\mu \mathcal{H}'^{-}_{1}$ & $ig s_W$ \\
$A_\mu \mathcal{H}'^{+}_{2} \overlr{\partial}^\mu \mathcal{H}'^{-}_{2}$ & $ig s_W$ & $Z_\mu  H_2 \overlr{\partial}^\mu \mathcal{A}_{1}$ & $\fr{-g}{2c_W}\Upsilon s_{\ep_1}$ \\
$Z_\mu H_2 \overlr{\partial}^\mu \mathcal{A}_{2}$ & $\fr{g}{2c_W}\Upsilon c_{\ep_1}$ & $Z_\mu H_5^{0*} \overlr{\partial}^\mu H_5^0$ & $\fr{ig}{c_W}$ \\
$Z_\mu H_{11}^{++} \overlr{\partial}^\mu H_{11}^{--}$ & $\fr{ig}{c_W}c_{2W}$ & $Z_\mu H_{12}^- \overlr{\partial}^\mu H_{12}^+$ & $igs_Wt_W$ \\
$Z_\mu H_{1}^{+} \overlr{\partial}^\mu H_{1}^{-}$ & $\fr{ig}{2c_W}c_{2W}$ & $Z_\mu \mathcal{H}'^{+}_1 \overlr{\partial}^\mu \mathcal{H}'^{-}_1$ & $\fr{ig}{2c_W}(c_{2W}-c_{\ep_2}^2)$ \\
$Z_\mu \mathcal{H}'^{-}_1 \overlr{\partial}^\mu \mathcal{H}'^{+}_2$ & $\fr{ig}{4c_W}s_{2\ep_2}$ & $Z_\mu \mathcal{H}'^{-}_2 \overlr{\partial}^\mu \mathcal{H}'^{+}_2$ & $\fr{ig}{2c_W}(s_{\ep_2}^2-c_{2W})$ \\
\end{tabular}
\caption{\label{1NG2STab1}The interactions of a neutral gauge boson with two scalars.}
\end{ruledtabular}
\end{table}

\begin{table}[!h]
\begin{ruledtabular}
\begin{tabular}{lccc}
Vertex & Coupling & Vertex & Coupling\\ \hline
$W^+_\mu H_2 \overlr{\partial}^\mu H_1^-$ & $\fr{-ig}{2}$ & $W^+_\mu H_5^0 \overlr{\partial}^\mu H_{12}^-$ & $ig$ \\
$W^+_\mu H_{11}^{--} \overlr{\partial}^\mu H_{12}^+$ & $-ig$ & $W^+_\mu H_1^- \overlr{\partial}^\mu \mathcal{A}_1$ & $\fr{-g}{2}\Up s_{\ep_1}$ \\
$W^+_\mu H_1^- \overlr{\partial}^\mu \mathcal{A}_2$ & $\fr{g}{2}\Up c_{\ep_1}$ & $W^+_\mu \mathcal{H}'^-_1 \overlr{\partial}^\mu H'^0_1$ & $\fr{ig}{\sqrt2}(\fr v w c_{\al_4}c_{\ep_3}+s_{\ep_3})s_{\ep_2}$\\
$W^+_\mu \mathcal{H}'^-_1 \overlr{\partial}^\mu H'^0_2$ & $\fr{-ig}{\sqrt2}s_{\ep_2}c_{\ep_3}$ & $W^+_\mu \mathcal{H}'^-_2 \overlr{\partial}^\mu H'^0_1$ & $\fr{-ig}{\sqrt2}(\fr v w c_{\al_4}c_{\ep_3}+s_{\ep_3})c_{\ep_2}$\\
$W^+_\mu \mathcal{H}'^-_2 \overlr{\partial}^\mu H'^0_2$ & $\fr{ig}{\sqrt2}c_{\ep_2}c_{\ep_3}$ & $X^+_\mu H_1 \overlr{\partial}^\mu  \mathcal{H}'^-_1$ & $\fr{-ig}{2}c_{\al_1}c_{\ep_2}$ \\
$X^+_\mu H_1 \overlr{\partial}^\mu  \mathcal{H}'^-_2$ & $\fr{-ig}{2}c_{\al_1}s_{\ep_2}$ & $X^+_\mu H_2 \overlr{\partial}^\mu  \mathcal{H}'^-_1$ & $\fr{-ig}{2}s_{\al_1}c_{\ep_2}$\\
$X^+_\mu H_2 \overlr{\partial}^\mu  \mathcal{H}'^-_2$ & $\fr{-ig}{2}s_{\al_1}s_{\ep_2}$ & $X^+_\mu H_3 \overlr{\partial}^\mu  \mathcal{H}'^-_2$ & $\fr{ig}{2}(\sqrt2 s_{\al_2}s_{\al_4}+c_{\al_2}c_{\al_4})c_{\ep_2}$ \\
\end{tabular}
\caption{\label{1CG2STab1}The interactions of a charged gauge boson with two scalars.}
\end{ruledtabular}
\end{table}

\begin{table}[!h]
\begin{ruledtabular}
\begin{tabular}{lccc}
Vertex & Coupling & Vertex & Coupling\\ \hline
$H_1 Z Z'$ & $\fr{g^2}{2c_W^2\sqrt{1+2c_{2W}}}(uc_{\al_1}-vc_{2W}s_{\al_1})$ & $H_1 Z Z$ & $\fr{g^2}{4c_W^2}\sqrt{u^2+v^2}$ \\
 $H_1 Z' Z'$ & $\fr{g^2}{4c_W^2(1+2c_{2W})}(uc_{\al_1}+vc_{2W}^2s_{\al_1})$ & $H_2 Z Z'$ & $\fr{g^2}{\sqrt{1+2c_{2W}}}us_{\al_1}$\\
 $H_2 Z' Z'$ & $\fr{g^2}{1+2c_{2W}}us_W^2s_{\al_1}$ & $H_3 Z' Z'$ & $\fr{g^2}{1+2c_{2W}}c_W^2(wc_{\al_2}-4\La s_{\al_2})$\\
 $H_4 Z' Z'$ & $\fr{g^2}{1+2c_{2W}}c_W^2(ws_{\al_2}+4\La c_{\al_2})$ & $H'^0_1 Z Y^{0*}$ & $\fr{-g^2}{2c_W}vc_{\ep_3}$\\
 $H'^0_1 Z' Y^{0*}$ & $\fr{g^2}{2c_W^2\sqrt{3-t_W^2}}(2wc_W^2 c_{\al_4} s_{\ep_3}-v c_{\ep_3})$ & $H'^0_2 Z Y^{0*}$ & $\fr{-g^2}{4c_W}v(s_{\ep_3}+\fr v w c_{\al_4}c_{\ep_3})$\\
 $H'^0_2 Z' Y^{0*}$ & $\fr{-g^2}{4c_W^2\sqrt{3-t_W^2}}(4wc_W^2 c_{\al_4} c_{\ep_3}+v s_{\ep_3})
 $ & $H_1 W^+ W^-$ & $\frac{g^2}{2}\sqrt{u^2+v^2}$\\
 $H_1 X^+ X^-$ & $\frac{g^2}{2}uc_{\al_1}$ & $H_1 Y^0 Y^{0*}$ & $\frac{g^2}{2}vs_{\al_1}$\\
\end{tabular}
\caption{\label{1S2GTab1}The interactions of a scalar with two gauge bosons.}
\end{ruledtabular}
\end{table}

\begin{table}[!h]
\begin{ruledtabular}
\begin{tabular}{lccc}
Vertex & Coupling & Vertex & Coupling\\ \hline
$A W^+ H'^0_1 \mathcal{H}'^-_1$ & $\frac{-g^2}{\sqrt2}s_W(s_{\ep_3}+\fr v w c_{\al_4}c_{\ep_3})s_{\ep_2}$ & $A W^+ H'^0_1 \mathcal{H}'^-_2$ & $\frac{g^2}{\sqrt2}s_W(s_{\ep_3}+\fr v w c_{\al_4}c_{\ep_3})c_{\ep_2}$ \\
$A W^+ H'^0_2 \mathcal{H}'^-_1$ & $\frac{g^2}{\sqrt2}s_Ws_{\ep_2}c_{\ep_3}$ & $A W^+ H'^0_2 \mathcal{H}'^-_2$ & $\frac{-g^2}{\sqrt2}s_Wc_{\ep_2}c_{\ep_3}$ \\
$A W^+ H_{11}^{--} H_{12}^+$ & $3g^2s_W$ & $A W^+ H_{12}^-H_5^0$ & $g^2s_W$ \\
$A W^+ H_1^- H_2$ & $\fr{-g^2}{2}s_W$ & $A W^+ H_1^- \mathcal{A}_1$ & $\fr{-ig^2}{2}\Up s_Ws_{\ep_1}$ \\
$A W^+ H_1^- \mathcal{A}_2$ & $\fr{ig^2}{2}\Up s_Wc_{\ep_1}$ & $A X^+ H_{12}^- H'^0_1$ & $\fr{-g^2}{\sqrt2}s_Ws_{\al_4}s_{\ep_3}$ \\
$A X^+ H_{12}^- H'^0_2$ & $\fr{g^2}{\sqrt2}s_Ws_{\al_4}c_{\ep_3}$ & $A X^+ H_1^- H'^{0*}_1$ & $\fr{g^2}{\sqrt2}s_Wc_{\al_1}c_{\ep_3}$ \\
$A X^+ H_1^- H'^{0*}_2$ & $\fr{g^2}{\sqrt2}s_Wc_{\al_1}(s_{\ep_3}+\fr v w c_{\al_4}c_{\ep_3})$ & $A X^+ H_{11}^{--} \mathcal{H}'^+_1$ & $\frac{3g^2}{\sqrt{2\La^2+w^2}}s_W(ws_{\ep_2}+uc_{\al_4}c_{\ep_2})$ \\
\end{tabular}
\caption{\label{1NG1CG2STab1}The interactions of a neutral and a charged gauge boson with two scalars (Table continued).}
\end{ruledtabular}
\end{table}

\begin{table}[!h]
\begin{ruledtabular}
\begin{tabular}{lccc}
Vertex & Coupling & Vertex & Coupling\\ \hline
  $Z X^+ H^{--}_{11} \mathcal{H}'^{+}_{1}$ & $\fr{3g^2}{2\sqrt{2\La^2+w^2}c_W}c_{2W}(ws_{\ep_2}+uc_{\al_4}c_{\ep_2})$ & $Z X^+ H^{--}_{11} \mathcal{H}'^{+}_{2}$ & $\fr{-3g^2}{2c_W} c_{2W}s_{\al_4}c_{\ep_2}$\\
  $Z Y^0 H^{0*}_{5} H'^{0}_{1}$ & $\fr{3g^2}{2c_W} s_{\al_4}s_{\ep_3}$ & $Z Y^0 H^{0*}_{5} H'^{0}_{2}$ & $\fr{-3g^2}{2c_W} s_{\al_4}c_{\ep_3}$\\
  $Z Y^0 H'^{0*}_{1} H_1$ & $\fr{-g^2}{4c_W} s_{\al_1}c_{\ep_3}$ & $Z Y^0 H'^{0*}_{1} H_2$ & $\fr{g^2}{4c_W} c_{\al_1}c_{\ep_3}$\\
 $Z Y^0 H'^{0*}_{1} H_3$ & $\fr{-g^2}{4c_W}[(\sqrt2 s_{\al_2}s_{\al_4}+c_{\al_2}c_{\al_4})s_{\ep_3}+\fr v w c_{\al_2}c_{\ep_3}]$ & $Z Y^0 H'^{0*}_2 H_1$ & $\fr{-g^2}{4c_W}s_{\al_1}(\fr v w c_{\al_4}c_{\ep_3}+s_{\ep_3}) $\\
 $Z Y^0 H'^{0*}_{1} H_4$ & $\fr{g^2}{4c_W}[(\sqrt2 c_{\al_2}s_{\al_4}-s_{\al_2}c_{\al_4})s_{\ep_3}-\fr v w s_{\al_2}c_{\ep_3}]$ & $Z Y^0 H'^{0*}_2 H_2$ & $\fr{g^2}{4c_W}c_{\al_1}(\fr v w c_{\al_4}c_{\ep_3}+s_{\ep_3}) $\\
 $Z Y^0 H'^{0*}_2 H_3$ & $\fr{g^2}{4c_W}(c_{\al_2}c_{\al_4}+\sqrt2 s_{\al_2}s_{\al_4})c_{\ep_3}$ & $Z Y^0 H'^{0*}_2 H_4$ & $\fr{g^2}{4c_W}(s_{\al_2}c_{\al_4}-\sqrt2 c_{\al_2}s_{\al_4})c_{\ep_3}$\\
 $Z Y^0 H^{+}_1 \mathcal{H}'^-_1$ & $\fr{g^2}{2\sqrt2 c_W}(1-2c_{2W})s_{\al_1}c_{\ep_2}$ &  $Z Y^0 H^{+}_1 \mathcal{H}'^-_2$ & $\fr{g^2}{2\sqrt2 c_W}(1-2c_{2W})s_{\al_1}s_{\ep_2}$\\
\end{tabular}
\caption{\label{1NG1CG2STab2}The interactions of a neutral and a charged gauge boson with two scalars (Continued).}
\end{ruledtabular}
\end{table}

\begin{table}
\begin{ruledtabular}
\begin{tabular}{lccc}
Vertex & Coupling & Vertex & Coupling\\ \hline
  $Z' X^+ \mathcal{H}'^-_{2} H_3$ & $\frac{-g^2}{4\sqrt{3-t_W^2}c_W^2}[c_{\al_2}c_{\al_4}+\sqrt2(4c_W^2+1) s_{\al_2}s_{\al_4}]c_{\ep_2}$ & $Z' X^+ H^-_{12} H'^0_2$ & $\frac{g^2}{\sqrt{8(3-t_W^2)}}(1-3t_W^2)s_{\al_4}c_{\ep_3}$\\
   $Z' X^+ \mathcal{H}'^-_{2} H_4$ & $\frac{-g^2}{4\sqrt{3-t_W^2}c_W^2}[s_{\al_2}c_{\al_4}-\sqrt2(4c_W^2+1) c_{\al_2}s_{\al_4}]c_{\ep_2}$ & $Z' X^+ H^{--}_{11} \mathcal{H}'^+_2$ & $\frac{g^2}{2\sqrt{3-t_W^2}}(3t_W^2-1)s_{\al_4}c_{\ep_2}$\\
   $Z' X^+ H^{--}_{11} \mathcal{H}'^+_1$ & $\frac{g^2}{2\sqrt{3-t_W^2}}(1-3t_W^2)(\fr{u}{\sqrt{2\La^2+w^2}}c_{\al_4}c_{\ep_2}+s_{\al_4}s_{\ep_2})$ & $Z' Y^0 H'^{0*}_{1} H_1$ & $\frac{-g^2}{4c_W^2\sqrt{3-t_W^2}}s_{\al_1}c_{\ep_3}$\\
 $Z' Y^0 H'^{0*}_{2} H_1$ & $\frac{-g^2}{4c_W^2\sqrt{3-t_W^2}}s_{\al_1}(\fr v w c_{\al_4}c_{\ep_3}+s_{\ep_3})$ & $Z' Y^0 H'^{0*}_{1} H_2$ & $\frac{g^2}{4c_W^2\sqrt{3-t_W^2}}c_{\al_1}c_{\ep_3}$\\
   $Z' Y^0 H'^{0*}_{2} H_2$ & $\frac{g^2}{4c_W^2\sqrt{3-t_W^2}}c_{\al_1}(\fr v w c_{\al_4}c_{\ep_3}+s_{\ep_3})$ & $Z' Y^0 H^{+}_{1} \mathcal{H}'^-_1$ & $\frac{g^2}{\sqrt{8(3-t_W^2)}}(3t_W^2-1)s_{\al_1}c_{\ep_2}$ \\
  $Z' Y^0 H'^{0*}_{2} H_3$ & $\frac{g^2}{4c_W^2\sqrt{3-t_W^2}}[c_{\al_2}c_{\al_4}+\sqrt2(4c_W^2+1) s_{\al_2}s_{\al_4}]c_{\ep_3}$ &  $Z' Y^0 H^{+}_{1} \mathcal{H}'^-_2$ & $\frac{g^2}{\sqrt{8(3-t_W^2)}}(3t_W^2-1)s_{\al_1}s_{\ep_2}$\\
   $Z' Y^0 H'^{0*}_{2} H_4$ & $\frac{g^2}{4c_W^2\sqrt{3-t_W^2}}[s_{\al_2}c_{\al_4}-\sqrt2(4c_W^2+1) c_{\al_2}s_{\al_4}]c_{\ep_3}$ & $Z' Y^0 H^{0*}_{5} H'^{0}_1$ & $\frac{g^2}{\sqrt{4(3-t_W^2)}}(3t_W^2-1)s_{\al_4}s_{\ep_3}$\\
\end{tabular}
\caption{\label{1NG1CG2STab3}The interactions of a neutral and a charged gauge boson with two scalars (Continued).}
\end{ruledtabular}
\end{table}

\begin{table}[!h]
\begin{ruledtabular}
\begin{tabular}{lccc}
Vertex & Coupling & Vertex & Coupling\\ \hline
$W^+W^+ H^{--}_{11} H^0_5$ & $g^2$ & $W^+W^- H_1 H_1$ & $\frac{g^2}{4} $ \\
$W^+W^- H^{++}_{11} H^{--}_{11}$ & $g^2$ & $W^+W^- H^+_{12} H^-_{12}$ & $2g^2$ \\
$W^+W^- H^{+}_{1} H^{-}_{1}$ & $\fr{g^2}{2}$ & $W^+W^- H_2 H_2$ & $\fr{g^2}{4}$ \\
$W^+W^- \mathcal{H}'^{+}_{1} \mathcal{H}'^{-}_{1}$ & $\fr{g^2}{2}s_{\ep_2}^2$ & $W^+W^- \mathcal{H}'^{+}_{1} \mathcal{H}'^{-}_{2}$ & $\fr{-g^2}{4}s_{2\ep_2}$ \\
$W^+W^- \mathcal{H}'^{+}_{2} \mathcal{H}'^{-}_{2}$ & $\fr{g^2}{2}c_{\ep_2}^2$ & $W^+W^- H^{0}_{5} H^{0*}_{5}$ & $g^2$ \\
$W^+W^- H'^{0}_{1} H'^{0*}_{1}$ & $\fr{g^2}{2}(s_{\ep_3}^2+\fr {v}{w} c_{\al_4}s_{2\ep_3})$ & $W^+W^- H'^{0}_{1} H'^{0*}_{2}$ & $\fr{-g^2}{2}(s_{\ep_3}+\fr v w c_{\al_4}c_{\ep_3})c_{\ep_3}$ \\
$W^+W^- H'^{0}_{2} H'^{0*}_{2}$ & $\fr{g^2}{2}c_{\ep_3}^2$ & $W^+W^- \mathcal{A}_1 \mathcal{A}_1$ & $\fr{g^2}{4}\Up^2 s_{\ep_1}^2$ \\
\end{tabular}
\caption{\label{2CG2STab1}The interactions of two charged gauge bosons with two scalars (Table continued).}
\end{ruledtabular}
\end{table}

\begin{table}[!h]
\begin{ruledtabular}
\begin{tabular}{lccc}
Vertex & Coupling & Vertex & Coupling\\ \hline
$X^+X^+ H^{--}_{11} \mathcal{A}_1$ & $\fr{-ig^2}{\sqrt2}(s_{\al_3}c_{\ep_1}+c_{\al_1}s_{\al_1}c_{\al_3}s_{\ep_1})$ & $X^+X^+ H^{--}_{11} H_3$ & $\fr{-g^2}{\sqrt2}s_{\al_2}$\\
$X^+X^+ H^{--}_{11} \mathcal{A}_2$ & $\fr{ig^2}{\sqrt2}(c_{\al_1}s_{\al_1}c_{\al_3}c_{\ep_1}-s_{\al_3}s_{\ep_1})$ & $X^+X^+ H^{--}_{11} H_4$ & $\fr{g^2}{\sqrt2}c_{\al_2}$ \\
$X^+X^- H_3 H_3$ & $\fr{g^2}{4}(1+s_{\al_2}^2)$ & $X^+X^- H_1 H_1$ & $\fr{g^2}{4}c_{\al_1}^2$ \\
$X^+X^- H_4 H_4$ & $\fr{g^2}{4}(1+c_{\al_2}^2)$ & $X^+X^- H_2 H_2$ & $\fr{g^2}{4}s_{\al_1}^2$ \\
$X^+X^- H_1 H_2$ & $\fr{g^2}{4}s_{2\al_1}$ & $X^+X^- H_3 H_4$ & $\fr{-g^2}{4}s_{2\al_2}$ \\
$X^+X^- H^{++}_{11} H^{--}_{11}$ & $g^2$ & $X^+X^- H^+_{12} H^-_{12}$ & $\fr{g^2}{2}$ \\
$X^+X^- H'^0_1 H'^{0*}_2$ & $\fr{g^2}{2}c_{\al_4}(c_{\al_4}s_{\ep_3}+\fr v w c_{\ep_3})c_{\ep_3}$ & $X^+X^- H'^0_1 H'^{0*}_1$ & $\fr{g^2}{2}(1-c_{\al_4}^2s_{\ep_3}^2)$ \\
\end{tabular}
\caption{\label{2CG2STab2}The interactions of two charged gauge bosons with two scalars (Continued).}
\end{ruledtabular}
\end{table}

\begin{table}[!h]
\begin{ruledtabular}
\begin{tabular}{lccc}
Vertex & Coupling & Vertex & Coupling\\ \hline
$A A H^{++}_{11} H^{--}_{11}$ & $4g^2 s_W^2$ & $A A H^{+}_{12} H^{-}_{12}$ & $g^2 s_W^2$ \\
$A A H^{+}_{1} H^{-}_{1}$ & $g^2 s_W^2$ & $A A \mathcal{H}'^{+}_{1} \mathcal{H}'^{-}_{1}$ & $g^2 s_W^2$ \\
$A A \mathcal{H}'^{+}_{2} \mathcal{H}'^{-}_{2}$ & $g^2 s_W^2$ & $A Z H^{+}_{12} H^{-}_{12}$ & $-2g^2 s_W^2t_W$ \\
$A Z \mathcal{H}'^{+}_{1} \mathcal{H}'^{-}_{1}$ & $g^2 t_W(c_{2W}-c_{\ep_2}^2)$ & $A Z \mathcal{H}'^{+}_{1} \mathcal{H}'^{-}_{2}$ & $-g^2t_Wc_{\ep_2}s_{\ep_2}$ \\
$A Z \mathcal{H}'^{+}_{2} \mathcal{H}'^{-}_{2}$ & $g^2t_W (c_{2W}-s_{\ep_2}^2)$ & $A Z H^{++}_{11} H^{--}_{11}$ & $4g^2c_{2W}t_W$ \\
$A Z H^{+}_{1} H^{-}_{1}$ & $\fr{g^2}{2c_{W}}(s_{3W}-s_W)$ & $A Z' H^{++}_{11} H^{--}_{11}$ &$ \frac{4g^2}{\sqrt{1+2c_{2W}}}c_{2W}t_W$\\
$A Z' H^{+}_{12} H^{-}_{12}$ &$ \frac{2g^2}{\sqrt{1+2c_{2W}}}c_{2W}t_W$ & $A Z' H^{+}_{1} H^{-}_{1}$ &$ \frac{g^2}{\sqrt{1+2c_{2W}}}(2c_{W}^2c_{\al_1}^2-1)t_W$\\
\end{tabular}
\caption{\label{2NG2STab1}The interactions of two neutral gauge bosons with two scalars (Table continued).}
\end{ruledtabular}
\end{table}

\begin{table}[!h]
\begin{ruledtabular}
\begin{tabular}{lccc}
Vertex & Coupling & Vertex & Coupling\\ \hline
$Z' Z' H^{+}_{1} H^{-}_{1}$ & $\fr{g^2}{4c_W^2(1+2c_{2W})}(1-s_{2W}^2c_{\al_1}^2)$ &  $Z' Z' \mathcal{H}'^{+}_{2} \mathcal{H}'^{-}_{2}$ & $\frac{g^2}{4c_W^4(3-t_W^2)}[s_{\al_4}^2c_{\ep_2}^2+c_{2W}^2(c_{\al_4}^2c_{\ep_2}^2+4s_{\ep_2}^2)]$\\
$Z' Z' H^{++}_{11} H^{--}_{11}$ & $\fr{g^2}{c_W^2(1+2c_{2W})}c_{2W}^2$ &  $Z' Z' H^{+}_{12} H^{-}_{12}$ & $\fr{g^2}{c_W^2(1+2c_{2W})}c_{2W}^2$\\
$Z' Z' H^{0}_{5} H^{0*}_{5}$ & $\fr{g^2}{c_W^2(1+2c_{2W})}c_{2W}^2$ & No data & No data\\
\hline
Vertex & \multicolumn{3}{c}{Coupling}\\ \hline
 $Z' Z' H'^{0}_{1} H'^{0*}_{1}$ & \multicolumn{3}{c}  {$\frac{g^2}{4c_W^4(3-t_W^2)}[\fr v w c_{2W}^2c_{\al_4}s_{2\ep_3}+4c_W^4c_{\ep_3}^2+(1 -s_{2W}^2c_{\al_4}^2)s_{\ep_3}^2]$}\\
 $Z' Z' H'^{0}_{1} H'^{0*}_{2}$ & \multicolumn{3}{c}  {$\frac{g^2}{4c_W^4(3-t_W^2)}[(4c_{W}^2-1)(\fr{v}{w} c_{\al_4}c_{\ep_3} +s_{\ep_3})-s_{2W}^2s_{\al_4}^2s_{\ep_3}]c_{\ep_3}$}\\
  $Z' Z' H'^{0}_{2} H'^{0*}_{2}$ & \multicolumn{3}{c}  {$\frac{g^2}{3-t_W^2}[1+\fr{v}{w} c_{\al_4}s_{2\ep_3}-s_{\al_4}^2c_{\ep_3}^2-\fr{1}{4c_W^4}( 4c_{W}^2c_{\al_4}^2-1)c_{\ep_3}^2]$}\\
  $Z' Z' \mathcal{H}'^{+}_{1} \mathcal{H}'^{-}_{1}$ & \multicolumn{3}{c}  {$\frac{g^2}{4c_W^4(3-t_W^2)}[\fr{u}{\sqrt{2\La^2+w^2}}s_{2W}^2s_{\al_4}c_{\al_4}s_{2\ep_2}+4c_{2W}^2c_{\ep_2}^2+ (1-s_{2W}^2c_{\al_4}^2)s_{\ep_2}^2]$}\\
   $Z' Z' \mathcal{H}'^{+}_{1} \mathcal{H}'^{-}_{2}$ & \multicolumn{3}{c}  {$\frac{g^2}{8c_W^4(t_W^2-3)}\{\fr{u}{\sqrt{2\La^2+w^2}}s_{2W}^2s_{2\al_4}c_{\ep_2}^2+[s_{2W}^2(4-c_{\al_4}^2)-3]s_{2\ep_2}\}$}\\
\end{tabular}
\caption{\label{2NG2STab2}The interactions of two neutral gauge bosons with two scalars (Continued).}
\end{ruledtabular}
\end{table}

\bibliographystyle{JHEP}

\bibliography{combine}

\end{document}